\documentstyle[12pt,epsf,aps,rotate]{revtex}
\begin{document}
\title{
\vspace*{-3cm}
\begin{flushright}
{\small
HUB-EP 97/96, 
KANAZAWA 97-23 \\
to appear in PRD \\
}
\end{flushright}
\vspace*{0.5cm}
Action and topological density carried by Abelian monopoles
in finite temperature pure $SU(2)$ gauge theory:\\
an analysis using RG smoothing
\thanks{Supported by
a Visitorship of S.~T. to the Graduiertenkolleg
'Structure, Precision Tests and Extensions
of the Standard Model of Elementary Particle Physics', GRK 271/1-96}
\vspace*{1.0cm}
}
\author{
E.--M.~Ilgenfritz $^{1,2}$
\thanks{Supported by the Deutsche Forschungsgemeinschaft under grant Mu932/1-4},
H.~Markum $^3$,
M.~M\"uller--Preussker $^1$, 
and S.~Thurner $^{1,3}$ \\
$ $ \\
$^1$ {\it Institut f\"ur Physik, Humboldt--Universit\"at zu Berlin, Germany} \\
$^2$ {\it Institute for Theoretical Physics, University of Kanazawa, Japan} \\
$^3$ {\it Institut f\"ur Kernphysik, Technische Universit\"at Wien, Austria}\\
}
\maketitle
\begin{abstract}
\par\vspace*{2.0cm}\noindent
We test a new parametrization of a suitably truncated classically perfect
action for $SU(2)$ pure gauge theory 
with respect to self-consistency 
and locate the deconfinement transition on a $12^3\times4$ lattice. 
Using the technique of smoothing (blocking followed by inverse blocking)
we demonstrate clustering of action and topological charge density.
Concentrating on the Abelian monopoles found in smoothed configurations
after Abelian projection from the maximally Abelian gauge, we present
evidence for their role as carriers of non-Abelian action and 
topological charge.

\vspace{1cm}

\end{abstract}
\newpage

\section{Introduction}

Already for more than two decades the vacuum state as
well as the finite-temperature structure of QCD
have been viewed in terms of different kinds of topological
excitations, instantons and Abelian monopoles. On one hand,
instantons explain chiral symmetry breaking, play an important role
within the axial anomaly and more generally, they provide a successful
model for a wide range of applications in non-perturbative QCD
phenomenology (for references see \cite{thooft1,CDG,dyak,Shuryak}).
However, as long as (anti-) instantons are considered to form a dilute
medium, they are hardly explaining the confinement of quarks. On the other
hand, confinement can be reasonably based on a dual superconductor scenario
invented by 't Hooft and Mandelstam \cite{tHooft81,Mandelstam76}. Similarly
as Cooper pairs form a condensate in a real superconductor,
Abelian (anti-) monopoles occurring in pure Yang-Mills theories as
singularities of an appropriate gauge condition are thought to condense
and to lead to confinement through a color-electric dual Meissner effect.
Lattice computations carried out by different groups
\cite{Kronfeld,Suzuki,we} provided increasing evidence that, at
least in the maximally Abelian gauge, the projected $U(1)$
degrees of freedom are the carriers of the long-range physics and that
monopoles really seem to condense \cite{diGiacomo,Polikarpov}.

For a long time, these pictures of the QCD vacuum were not linked
to each other. Only very recently it became more and more clear, that
instantons, monopoles and dyons, respectively, are closely related
\cite{borschier,wien,Brower}. The correlation becomes
especially clear, when - after minimizing the classical action of
previously Monte Carlo generated quantum gauge fields - (approximate)
classical field configurations are approached.

In this paper we are going to investigate this interrelation in more
detail within pure $SU(2)$ lattice gauge theory at finite temperature.
We will apply a lattice formulation with fixed-point action which was
invented in order to keep the lattice theory as close as possible to
continuum physics. One virtue of the fixed-point action approach is that
it has made it possible to treat topological excitations like instantons
in a proper way \cite{bern1,bern2,boulder}. 
Moreover, such formulations are based
on renormalization group blocking and inverse blocking transformations.
This allows to give a precise definition of the scale of ultraviolet
fluctuations which are integrated out from the
gauge fields and how to change this scale \cite{boulder,invblock1}.
For inverse blocking this statement holds as long as 
only one inverse blocking step is performed. 

In a recent paper the present authors, together with M. Feurstein
\cite{weNPB}, proposed to use the inverse blocking method as a general
tool suppressing ultraviolet fluctuations, in particular in order
to expose the medium- and long-range physics in generic non-Abelian
lattice gauge field configurations.

The concept of inverse blocking originally arose within the context of
a classically perfect lattice action and has been essential
for the construction of the perfect action or suitably truncated actions
of less complexity \cite{invblock1}.
The concept was then used to handle the otherwise difficult problem
of defining {\it the} topological charge of a given 
lattice configuration \cite{boulder}.
The actual measurement of topological charge was based on
a geometric method.

In the meantime, the method has been refined by repeating several steps
of subsequent blocking (superimposing each time different coarse lattices)
and inverse blocking ('smoothing cycles') \cite{boulder97}
with the aim to resolve the instanton structure of individual lattice
configurations. Doing that, strictly speaking, these authors gave up
the original idea of inverse blocking, which was to
preserve the structure of a given lattice configuration at any
scale above the upper blocking level.
Applying smoothing cycles - in a way very similar to the
standard minimization of the gauge field action ('cooling')
- does not only wipe out short-range fluctuations of
perturbative origin but destroys also small-size non-perturbative structures
(e.g. small instantons and distorted instantons in the real vacuum)
that one might want to study. Not unexpectedly, the gauge fields are driven
into configurations which exhibit
well-separated lumps of action and topology. Near to their
centers, these lumps can be well interpreted as single instantons
frozen out of the lattice configurations. But there remained
a puzzle 
\cite{boulder97}. 
Similar to moderate cooling, 
sufficiently large Wilson loops evaluated on the
smoothed lattice gauge field configurations were still showing an area
law with non-vanishing string tension. However, if
one mocks up the gauge fields as dilute instanton gas 
configurations, taking the positions and sizes from the frozen out
instantons, one fails to produce confinement. 
Something seems to be lost.

The original Refs. \cite{boulder} have been our starting point to look 
into more details of the description of lattice topology going beyond the
measurement of the total topological charge.
As in Ref. \cite{weNPB}, we prefer in our present work
the original point of view that one should preserve the structure of
the gauge fields - as it is given in the quantized vacuum above the
blocked level -  by doing just {\it one} smoothing step.

 From the very beginning, instead of studying the distribution of
topological charge alone, we have been also interested in the monopole
degrees of freedom and their interrelation with the topological
density \cite{weNPB}. 
In that paper we have
found that by the minimal smoothing procedure the noise of the topological
density as well as the number of Abelian monopole currents become 
considerably reduced while the string tension changes only insignificantly.

This is challenging since it seems to indicate
that the bulk of the monopole activity is {\it not} related to the property
of confinement. For instance, in the result of smoothing 
(after removal of most of the abundant monopoles) the  
finite temperature anisotropy of the remaining monopole currents
becomes enhanced and behaves qualitatively like an order
parameter of the confinement/deconfinement transition.
In the deconfined phase, only very few space-like monopole currents remain 
on the smoothed lattice configurations.

In Sections II and III of the present work we prove the stability of
a new parametrization of the fixed-point gauge action and determine
the coupling $\beta_{crit}$ where the deconfinement transition 
occurs for our $N_{\tau}=4$ lattice. 
In the following Sections IV and V we provide
additional evidence that a gauge invariant characterization of the
Abelian monopole degrees of freedom is possible. We point out local
correlations between the monopole currents of the maximally Abelian gauge and
the local action and the local topological density. These signals are
enhanced for smoothed configurations. Finally, we conclude and propose 
directions for further investigations in Section VI.

\section{Tests for Optimizing the Action}

The simplified fixed-point action \cite{boulder} has been
improved recently, and a suitably truncated version \cite{kovacs_private}
has been obtained. Again, it is parametrized in terms of only two types
of Wilson loops, plaquettes $U_{C_{1}}=U_{x,\mu,\nu}$ (type $C_{1}$)
and tilted $3$-dimensional $6$-link loops (type $C_{2}$) of the form
\begin{equation}\label{sixlinks}
U_{C_{2}}=U_{x,\mu,\nu,\lambda}=
U_{x,\mu}
U_{x+\hat{\mu},\nu}
U_{x+\hat{\mu}+\hat{\nu},\lambda}
U_{x+\hat{\nu}+\hat{\lambda},\mu}^+
U_{x+\hat{\lambda},\nu}^+
U_{x,\lambda}^+ \, \, ,
\end{equation}
and contains several powers of the linear action terms
corresponding to each loop of both types that can be drawn on the
lattice
\begin{equation}\label{eq:fp_action}
S_{FP}(U)=\sum_{type~ i} \sum_{C_{i}} \sum_{j=1}^{4}
w(i,j) (1 - {1 \over 2}~\mbox{Tr}~U_{C_{i}} )^j  \, \, .
\end{equation}
The parameters of this action have changed \cite{kovacs_private}
compared to \cite{boulder}
and are reproduced in Table \ref{tab:weights}.

Suppose we have a fine lattice of links $U$ and a coarse lattice of
links $V$ covering the same physical volume.
Being a classically perfect action, $S_{FP}$ can be evaluated on both
lattices and must satisfy the following
condition
\begin{equation}\label{eq:equality}
~~~~   \mbox{Min}_{U} \left(S_{FP}(U) 
       + \kappa ~ T(U,V)\right) = S_{FP}(V)
\, \, ,
\end{equation}
configuration by configuration over a representative ensemble
of equilibrium gauge field configurations $V$,
near to the continuum limit.
In other words,
the minimum of the expression in brackets on the left hand side
must saturate the lower bound
provided by the right hand side.
$T(U,V)$ is a certain non-negative functional related to the blockspin
transformation (see \cite{boulder,weNPB}).
Blocking means a mapping 
from the link fields $U$ given on the fine 
lattice on link variables $V$ which ought describe the same configuration 
on the next coarser lattice. This mapping is realized by the following
construction: 
\begin{eqnarray}\label{eq:blocking}
\tilde{V}_{x,\mu} & = & c_{1}^{block} ~ 
U_{x,\mu} U_{x+\hat{\mu},\mu}
      \nonumber \\
 & & + \sum_{\nu \neq \mu} c_{2}^{block} \left(
U_{x,\nu} U_{x+\hat{\nu},\mu} U_{x+\hat{\nu}+\hat{\mu},\mu}
U_{x+2\hat{\mu},\nu}^{+}  \right.  \\
 & &  ~ ~ ~ ~ ~ ~ ~ ~ +  \left. 
U_{x-\hat{\nu},\nu}^{+} U_{x-\hat{\nu},\mu}
U_{x-\hat{\nu}+\hat{\mu},\mu}
U_{x-\hat{\nu}+2\hat{\mu},\nu} \right) \, \, .   \nonumber
\end{eqnarray}
The coarse link variable $V_{x,\mu}$ has to be an 
$SU(2)$ group element which requires normalization
\begin{equation}\label{eq:normalization}
V_{x,\mu} = {{\tilde{V}_{x,\mu}} \over
       {\sqrt{{\mathrm det}\left(\tilde{V}_{x,\mu}\right)}}} \, \, .
\end{equation}
The blocking parameters 
\begin{eqnarray}\label{eq:blockparam}
c_{2}^{block} & = & 0.12   \nonumber  \\
c_{1}^{block} & = & 1 - 6~ c_{2}^{block}
\end{eqnarray}
have been optimized earlier in Ref. \cite{bern1} under very 
general conditions.

Inverse blocking is  a mapping $V \to U^{SM}$ where the new link variables
$U = U^{SM}$ are implicitly defined as the set of links providing,
for the configuration encoded as $V$, 
the 'smoothest interpolation' 
on the next finer grained lattice. Practically, the link
field $U^{SM}$ must be found by minimizing the extended action in the
bracket on the left hand side of Eq. (\ref{eq:equality}). 
It is smoother than all possible {\it quantum} fields $U$ 
from which the coarse level field $V$ might have been obtained by
blocking. 
From the point of view of minimization, 
$\kappa$
plays the role of a Lagrange multiplier defining the strength of the
constraint provided by the coarse grained configuration. The relation
(\ref{eq:equality})
expresses the equality of the fine and coarse grained partition functions
within the saddle point approximation neglecting one-loop corrections
(which holds in the classical limit for $\beta \rightarrow \infty$).

In our last paper \cite{weNPB} we encountered
a problem that we could not resolve and which has caused some discussion.
On one hand we used the action proposed in Ref. \cite{boulder}. With this
action we have generated our Monte Carlo configurations
on the {\it fine lattice}. These have been blocked in order to
get the ensemble of coarse configurations $V$.
Then we did the minimization
to yield the respective smoothed copies $U^{SM}$ corresponding to each
Monte Carlo configuration $U^{MC}$, again on the fine lattice.
On the other hand, however, in order to match the condition of saturation
(\ref{eq:equality}) between the blocked and smoothed configurations
on average in our Monte
Carlo ensemble, we were forced to empirically tune $\kappa$ to 
another (smaller) effective value.
Thus, we were effectively working with a
weaker
constraint. Although we could make a number of valuable physical
observations, this was not satisfactory, because all parameters of the
extended action are strongly related to each other by the very
construction of this action.

Therefore, it was of primary importance for us to make an independent 
check of the new
parametrization of the perfect action 
using our best minimization procedure. 
Before presenting the results of this test
and the comparison with the old parametrization,
some details on our minimization routine used to accomplish the inverse
blocking are due at this place. There are two modes of relaxation which
differ in the way how a tentative update of the whole lattice is proposed.
They are tried one after another. 
The first mode consists in determining the direction of search for 
each link using the derivative with respect to $U$
of the {\it linear part} of the extended action 
$S_{FP}(U) + \kappa T(U,V)$ occurring on the left hand side
of Eq. (\ref{eq:equality}).
Then we try a quadratic interpolation of the extended 
action (now including the {\it non-linear
terms in} $S_{FP}$) as function of the given link along the corresponding 
geodesics in $SU(2)$. This interpolation is based on
action measurements for two 
step sizes (rotation angles) and is used to estimate 
an optimal angular step. This rotation is then applied to the link $U$
as a tentative update.
After all links are updated sequentially in this way,
a global check is made 
whether the {\it full extended action} 
(including now
the full blocking kernel $T(U,V)$)
would really decrease.
If this is the case, the tentative global update is accepted.
If not, the second relaxation mode is chosen. Now the direction of search 
is chosen randomly in order to explore the neighborhood
of the given link $U$ 
in the gauge group.  
With respect to
these directions a quadratic interpolation in the step size is 
constructed, too, in order to 
get an optimal rotation angle. 
If this global step leads to a smaller
value of the full extended action, the proposed update is accepted.
If not, the random search mode is repeated up to $5$ times. 
In the test runs, we made up to $15$ attempts in this mode. 
In most cases the first relaxation mode immediately
leads to a successful update.

The test of the two actions has been done for
$100$ Monte Carlo configurations on a $12^4$ lattice at $\beta=2.5$.
The two samples 
that we have to compare with respect to Eq. (\ref{eq:equality})
were obtained with the standard Metropolis algorithm 
after $1000$ thermalization steps, the configurations measured were 
separated by $100$ 
sweeps.
 The first one is obtained from the original Monte Carlo ensemble 
(created as $U^{MC}$) by blocking to $V(U^{MC})$.
The second sample is obtained from the first  
by relaxation (with $\kappa =12$) towards the smoothed $U^{SM}(U^{MC})$ fields.
We show in Fig. 1, 
for the new parametrization given in Table I,
the Monte Carlo history of the blocked action $S_{FP}(V)$
together with the respective minimum of the extended action 
$S_{FP}(U^{SM}) + \kappa T(U^{SM},V)$.
Average and standard deviation of the blocked action $S_{FP}(V)$ are 
$1988 \pm 25$
and for the minimum of the extended action  
$S_{FP}(U^{SM})+\kappa T(U^{SM},V)$ average and standard deviation are
$2024 \pm 30$. 
For the difference
$S_{FP}(U^{SM})+\kappa T(U^{SM},V) - S_{FP}(V)$, evaluated
for respective blocked and smoothed configurations, we find
$36 \pm 6$. 
These numbers give an impression
how well the smoothed configurations represent the
blocked ones. The old parametrization of the action has, for comparison
at the same $\beta=2.5$, average and standard deviation
$1509 \pm 25$  
for $S_{FP}(V)$,
$1869 \pm 28$ 
for the minimal extended action $S_{FP}(U^{SM})+\kappa T(U^{SM},V)$ and
$360 \pm 5$ 
for the difference. 
These results show to what extent the new action is better.
It is not completely perfect, however, as 
it cannot be so at any finite $\beta$.

To get a feeling how sensitive the coefficients of the action are
with respect to an attempted further improvement we considered the following.
The left hand side of (\ref{eq:equality}) depends on the parameters of
the action not only explicitly, but also through the selection
of Monte Carlo configurations $U^{MC}$ and through the result of relaxation,
$U^{SM}$. The dependence on the action coefficients is explicit only in
$S_{FP}(V)$.
Measuring the different contributions to the fixed-point action $S_{FP}$
for the configurations $U^{SM}$ and $V$ one can now try to optimize 
the - deliberately
free - parameters $\tilde{w}(i,j)$ in some modified action
$\tilde{S}_{FP}(V)$ by
minimizing the following error functional
\begin{equation}\label{eq:error}
~\sum_{{U^{MC}}} \left\{\tilde{S}_{FP}\left(V(U^{MC})\right) 
			   - S_{FP}\left(U^{SM}(U^{MC})\right) 
- \kappa ~ T\left(U^{SM}(U^{MC}),V(U^{MC})\right)\right\}^2
/ ~\sum_{{U^{MC}}} 1     \, \, .
\end{equation}
The average is over our test sample of configurations.
The  $\tilde{w}(i,j)$ are chosen initially at random near to $w(i,j)$ 
and then  they are let
to evolve to minimize the error functional (\ref{eq:error}).
This minimum turns out to be $\Delta S_{FP}=4.98$ in the case of the
new parametrization (applied to the fine lattice) and  $\Delta S_{FP}=6.21$
for the old one. More importantly, this requires small relative changes of the 
parameters $\tilde{w}(i,j)$ in the case 
of the new action, typically $\simeq 1$ percent, 6 percent for the worst 
critical coupling $w(2,1)$. 
For the old parametrization the relative 
changes vary from $O(10)$ percent up to $O(100)$ percent for the most 
critical $w(2,1)$ and $w(2,4)$.
In this sense, the new parametrization can be considered 
as much more stable.

In contrast to \cite{weNPB}, the theoretically favored value of $~\kappa=12~$
has been used in these tests
and throughout the further investigations reported in this paper. 
We emphasize again the 
improvement achieved with the new parametrization 
\cite{kovacs_private} compared to the old one
\cite{boulder}. 
In the present work 
the new parametrization of the fixed-point action
has been systematically tested and, 
to our knowledge for the first time, 
practically used for the topological analysis of equilibrium lattice
configurations.

\section{Search for the deconfinement transition}

We have simulated thermal configurations with the new parametrization of
the simplified fixed-point action at six  $\beta$ values
in the interval $[1.52,1.56]$ on 
lattice sizes $~L^3 \times 4~$ with $~L=8,10$ and $12$.
The statistics covered $1000$ 
configurations for the
$\beta$ values at the ends of the interval and $5000$ for the four  
interior $\beta$ values.
The measurements were always separated by $20$ Monte Carlo sweeps
using the Metropolis algorithm.
We have checked that the autocorrelations for the observables considered 
in the following were sufficiently suppressed.  
We examined the intersection of
the Binder cumulants of the volume--averaged Polyakov loop $P$
\begin{equation}
B_4(L)=\frac{\langle P^4 \rangle}{\langle P^2 \rangle^2}-3 \, \,
\end{equation}
measured on 
the three
lattices, respectively.
For simplicity the result is shown only for $~L=8, 12~$
in Fig. 2, in the upper part
for the cumulant 
$B^{U}_4(L)$
expressed in terms of the Polyakov loop calculated
from the $U^{MC}$ ({\it not} from the smoothed configuration)
on the fine lattice {\it and} in the lower part 
the cumulant
$B^{V}_4(L)$
in terms of the Polyakov loop calculated from the blocked links $V$.
These 
(unblocked and blocked)
cumulants behave very similar 
which gives independent support for the block spin
transformation used. 
We find the crossings of the Binder cumulants 
for the 
three $L$ values 
with each other
to occur within the same interval 
$\beta_{crit} \in \left[ 1.535, 1.550 \right]$, 
both for the unblocked $B^{U}_4(L)$ and the blocked
$B^{V}_4(L)$.
The corresponding values of the Binder cumulants have been seen to vary 
roughly in between $~-1.3~$ and $~-1.6~$.
There is an 
approximate agreement of $B^{U}_4(L)$ and $B^{V}_4(L)$ at $\beta_{crit}$ 
with the value of 
$B_4(L)$ expected from the $3d$ Ising model 
(see \cite{fingberg}). 
This gives an indication that the finite temperature pure $SU(2)$ 
theory with the perfect
action could also belong to the usual 
three-dimensional $Z(2)$ spin model universality class. 

In other words, the ensemble of blocked configurations
analyzed by means of the
cumulant crossing criterion for the thermal phase transition
points to the same value of $\beta_{crit}$
as on the fine lattice where the 
actual simulations have been carried out. Notice that the
coarse lattice has only $N_{\tau}=2$ temporal lattice steps.
We expect that the finite temperature results of a simulation on such 
a coarse lattice would suffer from bigger discretization errors.
In any case, 
sizes of the fine lattice similar to that used in this study 
are certainly too small 
to draw any conclusions with respect to the 
finite size scaling behaviour of the model. 

Before entering a thorough study over a broad range of temperatures
including the immediate neighborhood of the
deconfinement temperature, we will content ourselves
in the present work
with explorations at two temperatures, one in the confinement region
corresponding to $\beta=1.4$ and one 
deeply in the deconfinement phase at $\beta=1.8$.
We are aware that these cases are relatively far from the actual
transition temperature. The linear extension of the lattice at
$\beta=1.8$ changes roughly by a factor one half compared with 
the lower $\beta$ value.

\section{Clustering of action and topological charge
in smoothed configurations}

In this Section we will show that already
the procedure of minimal smoothing
(one blocking followed by one inverse blocking step)
exhibits the clustering of topological charge and action. 
On the $12^3\times4$ lattice used for our finite temperature investigation,
at lower $\beta$ values compared to the tests
of the action,
the inverse
blocking does not reach that degree of saturation of Eq. (\ref{eq:equality})
as decribed above.
In comparison with Fig. 1 the two histories are
somewhat more displaced from each other, and the remaining difference
amounts to a few percent of $S_{FP}(V)$ depending on $\beta$.

The full topological analysis was done for only $50$ configurations at 
both temperatures. 
We define the action density $s_{site}(x)$ per lattice point 
(also called 'local action') as follows
\begin{equation} 
\label{eq:local_action}
s_{site}(x) =  \sum_{j=1}^{4} 
\left(
  \sum_{C_{1}(x)}  \frac{w(1,j)}{4} (1 - {1 \over 2}~\mbox{Tr}~U_{C_{1}(x)})^j    
+ \sum_{C_{2}(x)}  \frac{w(2,j)}{6} (1 - {1 \over 2}~\mbox{Tr}~U_{C_{2}(x)})^j 
\right)
 \,. 
\end{equation}
Here, according to the notation  in Section II, 
$~C_{j}(x)$ (with $j=1,2$) means loops of type $j$ running through
the lattice site
$x$. Summing up $s_{site}(x)$ with respect to $x$ yields the total
action.
The topological charge density definitions 
we are using are the naive one constructed
out of plaquettes around a site $x$
\begin{equation}\label{eq:q_naive}
q(x) = - \frac1{2^9 \pi^2} \sum_{\mu,\nu,\sigma,\rho=-4}^{+4}
\epsilon_{\mu\nu\sigma\rho}
{\mathrm tr} \left(U_{x,\mu,\nu} U_{x,\sigma,\rho}\right)
\end{equation}
or the contribution from a given hypercube 
according to L\"uscher's
definition of charge \cite{luescher}. 
We did not
attempt to improve the field theoretic definition. Both topological
densities used are known to behave regularly
for smooth configurations \cite{weNPB}.

The sequence of both (smoothed) charges is shown
for $\beta=1.4$ as an example for
the
confinement phase 
in Fig. 3 and 
subsequent configurations
seem to be reasonably decorrelated. 
The ensemble 
has a dispersion of topological
charges $\langle Q^2 \rangle = 9.04$ in case of the integer valued 
L\"uscher charge and 
$\langle Q^2 \rangle = 7.14$ for the naive topological
charge. 

Figs. 4  show, in a $12 \times4$ array of sub-pictures, each
depicting $12\times12$ lattice points forming an $xy$ plane, 
the clustering of the topological
charge after smoothing. Figs. 4a, b correspond to a $~Q= 0~$ 
configuration obtained in the confinement phase.
The majority of configurations has 
total charge
$~Q \ne 0~$ but 
does not look qualitatively
different from the shown
example. The pattern of topological clusters is very rich. It is shown
here together with the few monopole world lines remaining after smoothing.
(Anti-) monopoles running in one of the $xy$ planes are shown by vertical and
horizontal arrows. Time-like (anti-) monopoles 
are symbolized by arrows to (from) the upper right, 
space-like (anti-) monopoles with world lines in the $z$ direction
by arrows to (from) the upper left. 

Figs. 4c, d  show in the same way a smoothed
configuration with total charge
$Q=0$ obtained at $\beta=1.8$ in the deconfinement
phase.
It illustrates that the 
still existing,
diluted 
topological charge is clustered
in more perfect instanton-like shape 
and 
how
the monopole world lines
are correlated with these clusters.

It is visible that our minimal smoothing method
gives clear evidence for clustering of the topological density
at a scale of several lattice spacings
in any possible section of the lattice
in both phases.
The topological density
appears as a sufficiently smooth function in all Euclidean coordinates.
However, a clear interpretation or parametrization in the form of instantons
seems hardly possible in the confinement phase.
Taking the action instead of the topological density, a similar behavior
is observed.

This means that our minimal variant of smoothing is not biased in
favor of nearly classical solutions of instanton shape. Nevertheless,
as Fig. 5 shows,
strong fields are preferably selfdual or antiselfdual. Physically, this
indicates that the (anti-) instantons entering the dense fluid 
become
strongly deformed
under the influence of quantum fluctuations.
A local analysis reveals that nearly selfdual or antiselfdual
domains are 
clearly 
preferred in both phases
as far as the local action exceeds a threshold value of
$s_{site}(x) \approx 0.3$. This is documented by the ridges in
the probability distribution of the local topological charge at
given local action density which is shown in
Fig. 5. 
It is not surprising that - at all $\beta$ values 
in a manner similar to the continuum -
the topological density is locally bounded by the action density
\begin{equation}
\frac{8 \pi^2}{g^2} ~|q(x)| \leq  \beta \cdot s_{site}(x) \, .
\end{equation}
More than that, we find an enhanced probability to find a topological density
near to saturation. This means that after smoothing
strong enough color-electric
and -magnetic fields show up mutually aligned to a high degree.

\section{Gauge invariant properties  of Abelian monopoles}

In our previous paper \cite{weNPB} we have performed the transformation to
the maximal Abelian gauge of smoothed configurations $U^{SM}$ 
and, for comparison, of the Monte Carlo configurations 
$U^{MC}$ they originated from. After the projection
to the Abelian gauge field (neglecting the charged vector matter fields
in the non-diagonal components of the links) \cite{Kronfeld}, the Abelian
monopole content of the original Monte Carlo ensemble and of the
ensemble of smoothed copies thereof
has been determined by the DeGrand--Toussaint construction
\cite{deGT} as usual.

Our most astonishing result was that the monopole activity was reduced by
more than an order of magnitude, roughly proportional to the reduction of the
topological activity. The latter has been defined in Ref. \cite{weNPB} as
$A_t=\sum_x |q(x)|$ in terms of the local topological density.
Although the correlation function of monopole currents and topological
density (given as a function of distance) was measurable also
without smoothing, it was not different in shape
(different in normalization, of course)
for Monte Carlo and smoothed configurations.

We give now more detailed evidence for the local correlation of gauge invariant
quantities (topological charge density and action density)
with the presence of monopole currents.
We have already 
illustrated the occurrence of (anti-) monopoles among the clusters
of topological charge by plotting their world lines on the dual lattice
in Figs. 4a, b in the confinement phase and in Figs. 4c, d in the
deconfinement phase. Now we want to make these observations more quantitative.
We did not consider here in detail the problem of non-uniqueness when fixing 
the maximal Abelian gauge (cf. \cite{we}).\footnote{Circumstancial evidence
shows that the monopole currents can be displaced by one lattice spacing
from one gauge copy to the other.}
Instead, we have 
driven
the corresponding gauge functional into a local extremum 
in the standard way. Our convergence criterium was chosen 
in such a way that we stopped the gauge fixing when  
the local relative changes of the gauge fixing functional become
globally less than $O(10^{-8})$. 
In the average for smoothed configurations, 
this has required $620~(690) \pm 250$
gauge cooling iterations for $\beta=1.4~(1.8)$. 
The number of gauge
cooling iterations necessary for original Monte Carlo configurations 
to reach the same accuracy was only slightly bigger.

Fig. 6 shows in its upper part how, in smoothed configurations,
the average occupation number of magnetic monopoles 
$\langle m(x) \rangle = \langle \sum_{dual~links~l~next~to~x} |m_l| \rangle$ 
on nearest dual links depends on the local action available.
There are $32$ dual links nearest to a given lattice point.
Here, the lattice points of all $50$ configurations obtained at
$\beta=1.4$ have been classified according to their local action.
For each bin the average number of monopole currents per lattice point
has been calculated.
The lower part of Fig. 6 shows the dependence of the average
number of monopole currents on the local topological charge density
for $\beta=1.4$.

Apart from obvious changes
in the scale of the action density,
generically
the same dependence occurs for 
$\beta=1.8$ in the deconfinement phase.
For instance, the range of the distribution of the
local action values varies with $\beta$.
In contrast to this, there are practically no lattice points with
topological density $|q(x)| > 0.05$, irrespective of $\beta$.
A physically relevant feature in Fig. 6
is the increase of the average local density of monopole currents
$\langle m(x) \rangle$
at small action density $s_{site}(x)$ or topological density $|q(x)|$.
For somewhat larger values of $s_{site}(x)$ and $|q(x)|$, respectively, 
the mean local monopole density seems to reach a plateau
of one monopole current on the average.
The increase of the error bars with $s_{site}$ or $|q|$ reflects the
insufficient statistics of lattice points with 
larger action or topological density.
 
Turning our point of view, we will now give a characterization
of Abelian monopole currents
in terms of locally defined {\it gauge independent} observables.
We consider the  $3$-cubes of the original lattice dual
to an elementary piece of monopole world line (a link of the dual lattice).
This leads us to a natural definition
of a local action on the monopole world line, $s_{3-cube}(c)$
instead of $s_{site}(x)$.
We include into $s_{3-cube}(c)$ the contribution of all plaquettes forming
the $6$ faces of the $3$-cube $c$ plus the contribution of all $6$-link
loops which wind around its surface. In short, it is exactly the part of the
total action which lives on that cube. Histograms of this
local cube--oriented action $s_{3-cube}(c)$ 
after smoothing 
are shown 
in Fig. 7, separately for time-like and
space-like dual links 
depending whether
they
are occupied by monopole currents or not.
Of course,
the range of values of $s_{3-cube}(c)$ spanned by these distributions depends
on $\beta$. 
All distributions are
normalized to one, such that the different numbers of cubes
entering each histogram
are not directly visible. 
In the confinement phase, the direction (time-like or space-like)
of the
dual link obviously does not matter.
The average of the local action and the square root of its variance 
$\langle \Delta s_{3-cube}(c)^2 \rangle$
are clearly bigger if the link carries a monopole current. 
This is qualitatively different in the deconfinement phase.
The rare space-like monopole currents are always associated with a
low local action
while the time-like
monopole currents have twice as large average 
action and bigger variance.

Fig. 8 analogously shows the histograms of a locally summed
topological charge, separately for time-like and
space-like dual links, occupied by a monopole current or not.
The sum over $q(x)$ includes here
the $8$ corners of the $3$-cube and should
capture a total charge inside of some domain around the cube
(centered in a hyperplane
orthogonal to the direction of the dual link).
The variance of the distribution is bigger for monopole cubes than for
empty cubes, and is independent of the direction of the monopole current
in the confinement phase. In the deconfinement phase, the variance of
the local topological charge is also bigger around cubes with monopoles.
Remember
that the histograms shown here are all normalized to one independent 
of the fact that there are only very few space-like monopoles (see below).

Smoothing is important to exhibit more clearly the action and topological
charge carried by the Abelian monopoles. In order to demonstrate this
by comparison, we present data for unsmoothed configurations in
Fig. 9,  only for one coupling at
$\beta=1.4$. Notice the order of magnitude change in the scale of the
local action $s_{3-cube}(c)$ compared to the plots for
smoothed configurations (Fig. 7).
Even without smoothing, there is a small but clear distinction
in the distribution of the cube--oriented local action
between
monopole cubes and cubes without monopoles.
In the 
topological density histograms, however,
one cannot see any difference between the presence and
absence of a monopole current as long as no smoothing has been performed.
This does not mean the absence of correlations between monopoles and 
topologically charged objects, it only reflects that the topological density
is hidden by quantum fluctuations.

It seems to be natural 
to define an
excess of action or topological charge per monopole world line length.
This should be possible (independent of the configuration) if monopoles
were semi-classical objects.
We define this as the mean of the local action $s_{3-cube}$
over all cubes which are dual to a monopole--occupied dual
link in a given configuration. In the case of the topological charge 
we use the mean  
of the geometric topological charge over all $4d$ hypercubes
from which a monopole current emanates.
Again, we study this separately for
space-like and time-like monopole currents. 

In Figs. 10 and 11 we compare the mean action and topological charge
per monopole world line length defined 
for the $50$ individual configurations in the
samples of
lattice configurations generated at $\beta=1.4$ and $\beta=1.8$, respectively.
The abscissa numbers the subsequent (smoothed) configurations.
The ordinate of the symbols represents the mean
local action or mean local 
topological charge per length of monopole 
world line in the configuration.
Circles refer to space-like and crosses to time-like pieces of monopole
world lines (monopole currents). 

Let us first look at the case of confinement (Fig. 10). There is
again no statistically significant difference between
time-like and space-like monopole currents. The mean local action 
and the mean local
topological charge per length
fluctuate strongly from configuration to configuration. 
This means that under the given circumstances there
is no sharp action or topological charge per length of world line 
measurable in the confinement phase.
The {\it ensemble} averages of these quantities coincide for
time-like and space-like currents within the variances.
The ensemble variances
and averages are considerably larger than the corresponding ones 
obtained for
the remaining cubes which are dual to empty links.

Secondly, let us consider the case of deconfinement (Fig. 11).
We have found only one configuration 
carrying a non-vanishing total topological charge. 
This is the configuration which stands out 
by the cross jumped up
in the lower left figure.
This configuration
has only time-like monopoles. One of the static monopoles is exactly
located at the center of the topological charge lump which
fluctuates but persists over
all Euclidean time slices. This object determines the total charge
$Q=1$. The opposite static antimonopole sits in a topologically non-active
part of the lattice.
In the high temperature ensemble, there are only four configurations 
in total which  
contain space-like monopole currents
(the circles in the left plots of the figure). 
Only for these configurations 
we can 
estimate
an action per length of space-like monopoles.
In spite of the low statistics, this 
points towards a
lower action
of space-like monopole currents than that of time-like monopoles
at this temperature, in accordance to Fig. 7.
Only for time-like monopole currents (mostly belonging to
perfectly static monopoles)
the averages of  
local action are markedly bigger than the respective quantities 
for cubes which are dual to empty
links. For the absolute value of the local topological charge the situation 
is not so clear.

The static monopoles in the deconfinement phase
are particle-like excitations possibly 
amenable to a semi-classical description.
On the contrary, discussing Fig. 10 we have seen that in the confinement phase
quantities 
like mean action or mean topological
charge per length of monopole world line fluctuate strongly
from configuration to configuration. 
Therefore, a semi-classical treatment 
of the Abelian monopoles seems to be questionable 
in the confinement phase.

Finally, in Fig. 12 the total length of monopole loops is shown, separated
into the time-like and space-like part, for each individual lattice field
configuration at $\beta=1.4$ and $\beta=1.8$, respectively.
In the confinement phase space-like monopole currents dominate over time-like
although there 
exists, at finite temperature, 
already some anisotropy (the ratio
is smaller than 
$~3$).
We find large clusters 
of monopole currents
which percolate through the lattice
both in time {\it and} in the
three space directions. This will be the topic of a
separate publication \cite{nextpaper}. 
In the deconfinement phase only
monopole-antimonopole pairs remain where each world line wraps
around the time direction.
In our sample, space-like currents occur exclusively as deformations of 
these {\it almost static} world lines, only
in the four configurations mentioned above.
This is clearly visible in Fig. 12 and
illustrates the direction in which the monopole content of the 
theory is changed in the
deconfinement phase. 
Nearer to the transition temperature 
we expect space-like monopole currents to occur more frequently but
no 
space-like percolation should exist above the transition temperature.

\section{Conclusions}

In this paper we have discussed in detail the topological structure
of pure $SU(2)$ lattice gauge theory at finite temperature. We
have analyzed lattice configurations simultaneously 
with respect to instanton-like topological 
excitations and with respect to 
Abelian monopoles (within the maximally Abelian gauge).
We have employed the framework of recently developed classically perfect
actions, more precisely a truncated fixed-point action.
Ultraviolet fluctuations were suppressed by techniques which play an
important
constructive role within the perfect action 
improvement program itself - blocking and inverse
blocking transformations  ('smoothing'). In contrast to various action
minimization schemes classified as  'cooling', smoothing represents just
one controllable constrained minimization step, which keeps the smoothed
fields close to the blocked fields at scale $2a$.

We have seen that the result of 'smoothing', nevertheless, 
looks quite similar to that of 'cooling',  if one restricts the latter
to very few and/or small iteration steps. The topological
structure becomes already clearly visible in terms of clusters of action and
topological charge ranging over several lattice spacings. These clusters
are certainly poorly described by 
ideal classical 
instanton solutions, but inside these clusters action and topological
charge vary continuously. At positions of high action density the smoothed
gauge fields turn out to be preferentially locally selfdual or anti-selfdual.

We have further investigated the correlation between instanton-like
excitations and Abelian monopoles which can be found after maximally 
Abelian gauge
fixing and Abelian projection. 
In order to illustrate this we have presented different views on the local
correlation by examining pieces of monopole world lines 
with respect to action and topological
charge densities attached to them. 

We found that despite their detection within
the maximally Abelian gauge, 
monopole currents alter the probability
distributions
of locally defined gauge independent quantities like action and 
topological density.
Monopoles
carry, in a statistical sense, significantly more gauge-invariant
quantities like action and topological charge, which might 
indicate that the dual superconductor mechanism has  
a gauge-invariant explanation. This is in the spirit of a recent 
investigation reported in Ref. \cite{bakker}. The
signal becomes enhanced for 
smoothed configurations. 

Independent of the {\it density} of monopole currents (being strongly 
reduced by smoothing) 
the relevant 
signal for monopole 
condensation is percolation in
all Euclidean directions. In particular, spatial percolation seems
to be needed to produce confinement in the usual sense, {\it i.e.} the
area law of
time-like Wilson loops. The monopole content also
of smoothed configurations fulfills this criterion in the 
confinement phase. Concerning deconfinement, we
have seen at large $~\beta~$ that exclusively almost static
monopole-like objects remain after smoothing.
They can
still explain 
magnetic screening
(the area law of space-like Wilson loops) in this phase.

For the near future we plan to 
investigate Abelian dominance after smoothing and to 
analyze the structure of the remaining dilute 
monopole currents in more detail.
We shall try to understand in as far the 
contribution of Abelian monopoles can quantitatively 
account for the area law of large Wilson loops.
We hope to be able to determine physically relevant scales 
like the average extension of
and the distance between instanton- and/or monopole-like clusters.

\par\vspace*{1cm}\noindent
{\bf Acknowledgments}

We are very indebted to the members of the Boulder group, 
in particular to T.~G.~Kovacs, who forwarded their new parametrization 
of the $SU(2)$ fixed-point action to us prior to publication.
S.~T. would like to thank M.~Feurstein for stimulating discussions.  
E.-M.~I., H.~M. and M.~M.-P. gratefully acknowledge the 
kind hospitality of T.~Suzuki and all other 
organizers of the {\it 1997 Yukawa International Seminar (YKIS'97)} 
on {\it Non-Perturbative QCD - Structure of the QCD Vacuum -} 
at the Yukawa Institute for Theoretical Physics of Kyoto University.


\newpage

\noindent
{\large TABLES: }

\vspace{1cm}

\begin{table}[h]
\begin{center}
\begin{tabular}{ l c c c c }
$w(i,j)$            & $j=1$   &  $j=2$  & $j=3$ & $j=4$   \\
\hline
$i=1$ (plaquettes)  & $1.115504$& $-.5424815$ & $.1845878$ & $-.01197482$ \\
$i=2$ (6-link loops)& $-.01443798$& $.1386238 $ & $-.07551325 $ & $.01579434$ 
\\
\end{tabular}
\end{center}
\caption{ Weight coefficients of the simplified fixed-point action}
\label{tab:weights}
\end{table}

\newpage

\noindent
{\large FIGURE CAPTIONS: }

\vspace{1cm}

\noindent
FIG. 1:
Monte Carlo histories of the 
blocked action $S_{FP}(V)$ and of the
extended action 
$S_{FP}(U^{SM})+\kappa T(U^{SM},V)$ 
for the smoothed configurations 
on a $12^4$ lattice at
$\beta=2.5$
as a test for the action to be used in
this work.\\

\noindent
FIG. 2:
Binder cumulants for the Polyakov loops before and after blocking. \\

\noindent
FIG. 3:
Monte Carlo sequence from the confinement phase at $\beta=1.4$
of topological charges in the 
smoothed configurations according to the naive and
L\"uscher's charge. \\

\noindent
FIG. 4:
(a) Clustering of topological density $q(x)$ 
after smoothing for 
a confinement configuration with $Q=0$ obtained
at $\beta=1.4$.
The subplots show $xy$-slices  for
$t=1, \cdots ,4$ and $z=1, \cdots ,6$.
Monopole currents are drawn as arrows. 
Time-like (anti-) monopoles 
are symbolized by arrows to (from) the upper right,
space-like (anti-) monopoles with world lines in the $z$ direction
by arrows to (from) the upper left.
(b) Same as (a) for $z=7, \cdots ,12$.
(c) Clustering of topological density $q(x)$ 
after smoothing for 
a deconfinement configuration with $Q=0$ obtained
at $\beta=1.8$.
The subplots show $xy$-slices  for
$t=1, \cdots ,4$ and $z=1, \cdots ,6$.
(d) Same as (c) for $z=7, \cdots ,12$. \\

\noindent
FIG. 5:
Probability distributions for finding a local topological
charge $q(x)$ at a lattice site 
together with a given local action $s_{site}(x)$.
$50$ independent confinement configurations obtained
at $\beta=1.4$
have been analyzed and the
$q$ distributions are separately normalized for all values of $s_{site}$. \\
\\

\noindent
FIG. 6:
Average occupation number of magnetic monopoles $\langle m \rangle$
on nearest dual links, 
depending on the local action  $s_{site}(x)$ (upper plot) and
depending on the topological density $q(x)$ (lower plot).
The figure depicts the situation in the confinement phase at $\beta=1.4$,
being similar to the deconfinement
phase. \\

\noindent
FIG. 7:
Probability distributions
for the action $s_{3-cube}(c)$ living on a $3$-cube
depending on whether the dual link is occupied ($m=1$) by a monopole
current or not ($m=0$). 
A distinction between time-like and space-like monopoles
is made.
$50$ independent configurations have been analyzed in both phases at
$\beta=1.4$ and $\beta=1.8$. 
The average values $\langle s_{3-cube}(c) \rangle$ 
and the square roots of the variances 
$\sqrt{\langle \Delta s_{3-cube}(c)^2 \rangle}$ are inserted. \\

\noindent
FIG. 8:
Probability distributions
for a local  topological charge (summed over the $8$ corners of a
$3$-cube) depending on whether the dual link is occupied ($m=1$) by a monopole
current or not ($m=0$).
The representation is the same as in the previous figure. \\

\noindent
FIG. 9:
Same as previous two figures but for unsmoothed configurations
at $\beta=1.4$ in the confinement phase.\\

\noindent
FIG. 10:
Mean local action $s_{3-cube}$ and Luescher's hypercube charge
for the sequence of Monte Carlo configurations 
depending on the monopole currents $|m|=1$ or $0$ on space-like
and time-like dual links
in the confinement
phase  at $\beta=1.4$.\\

\noindent
FIG. 11:
Same as previous figure in the deconfinement
phase at $\beta=1.8$. Note the isolated point in the lower left plot is due to 
$Q=1$. \\

\noindent
FIG. 12:
Monopole lengths 
for the sequence of Monte Carlo configurations 
in the same samples of $50$ independent
gauge field configurations for both phases
($\beta=1.4$ and
$\beta=1.8$)
as in the previous two figures. 
Distinction between time-like and space-like currents is made. \\

\newpage

\pagestyle{empty}
\textheight 25cm

\begin{figure}[!thb]
\label{fig:action_new}
\begin{center}
\begin{tabular}{c}
\\
\\
\\
\epsfxsize=11.0cm\epsffile{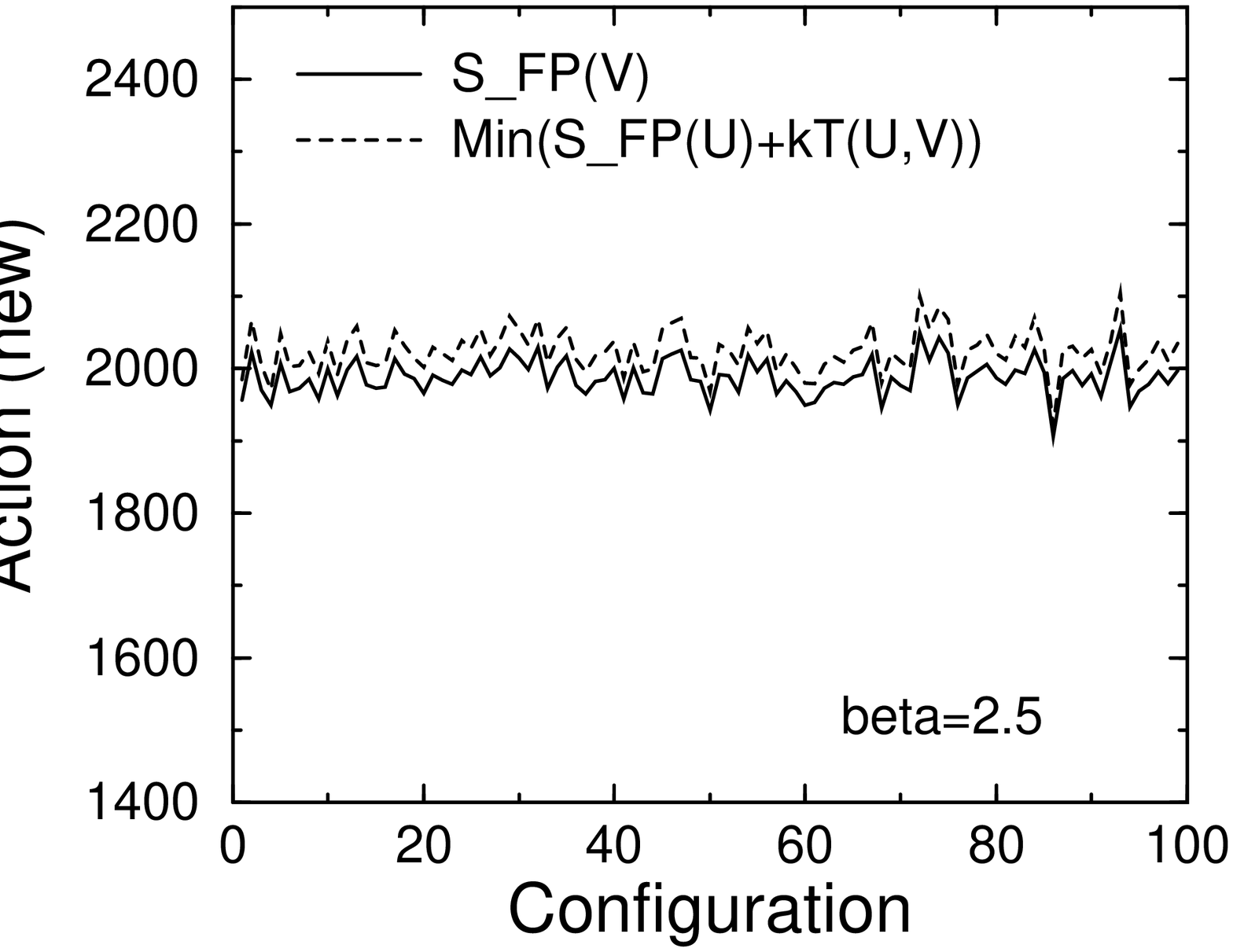}\\
\\
\\
\\
{FIG. 1 }
\end{tabular}
\end{center}
\end{figure}

\vspace{3cm}

\begin{figure}[!thb]
\label{fig:binder}
\begin{center}
\begin{tabular}{c}
\\
before blocking \\
\epsfxsize=10.0cm\epsffile{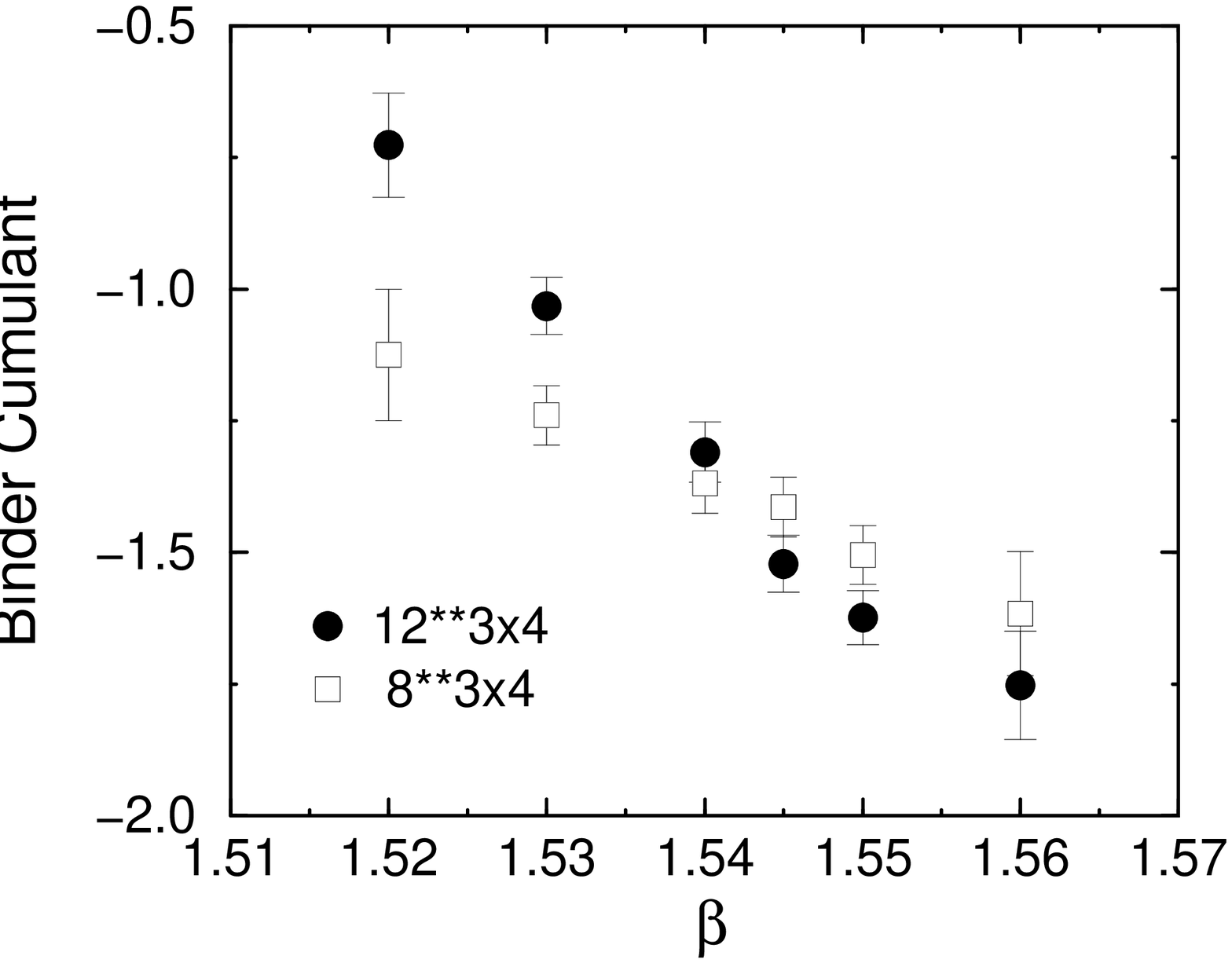}\\
\\
after blocking \\
\epsfxsize=10.0cm\epsffile{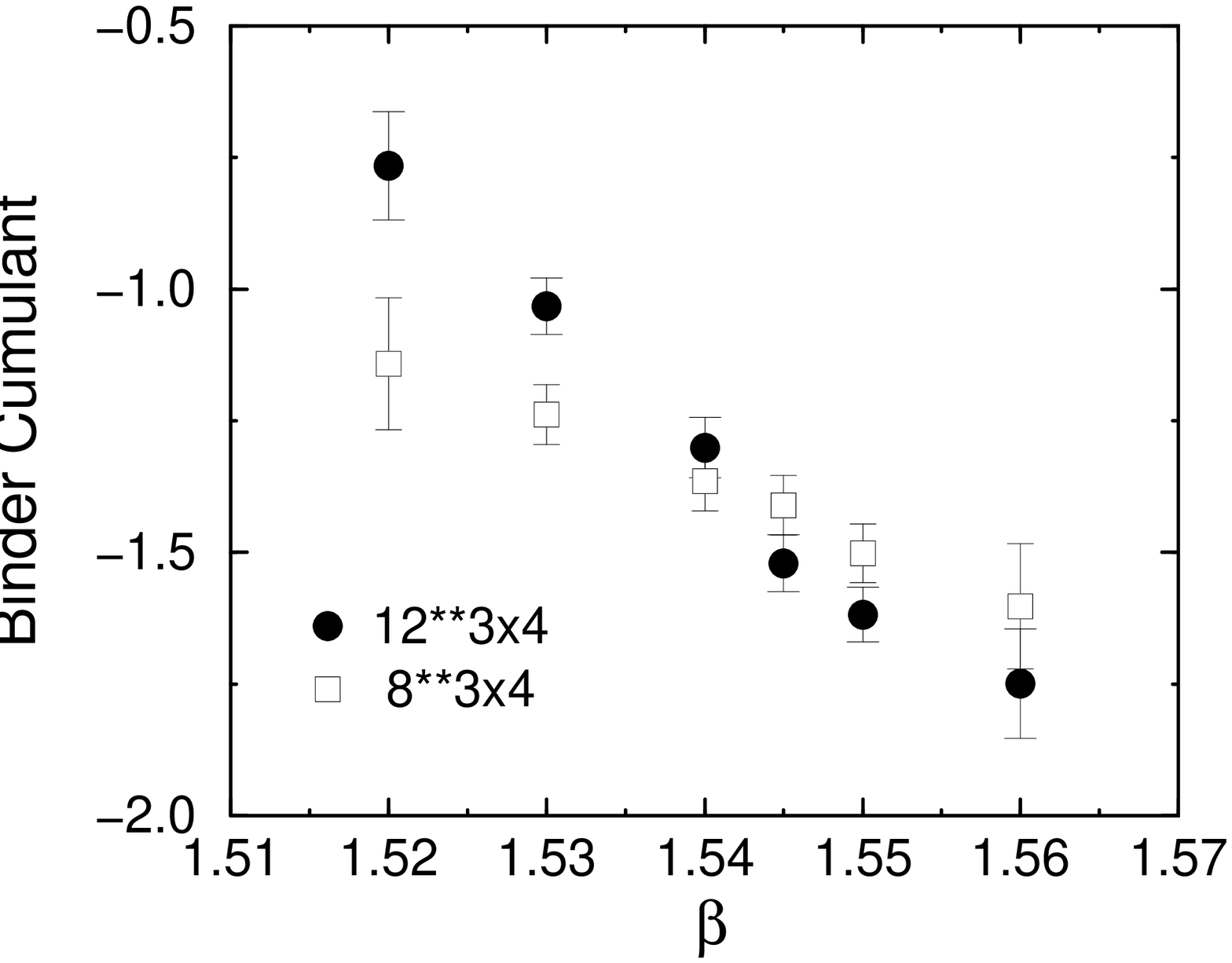}\\
\\
\\
\\
{FIG. 2 }
\end{tabular}
\end{center}
\end{figure}

\newpage

\begin{figure}[!thb]
\label{fig:hist140}
\begin{center}
\begin{tabular}{c}
\\
\\
\\
\epsfxsize=11.0cm\epsffile{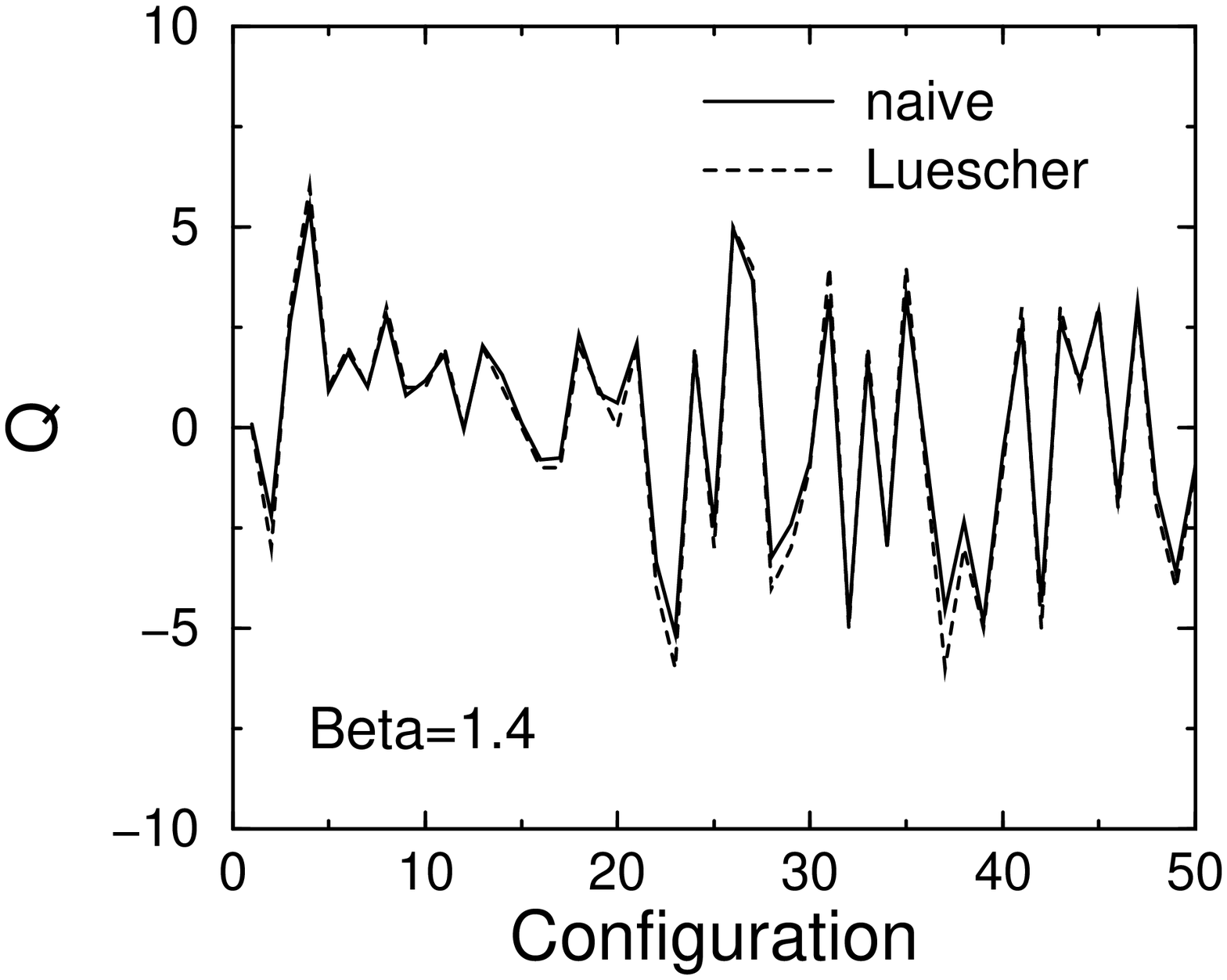}\\
\\
\\
\\
{FIG. 3 }
\end{tabular}
\end{center}
\end{figure}

\newpage

\begin{figure}[!thb]
\label{fig:Q-2beta1_1}
\begin{tabular}{c}
\epsfxsize=16.0cm\epsffile{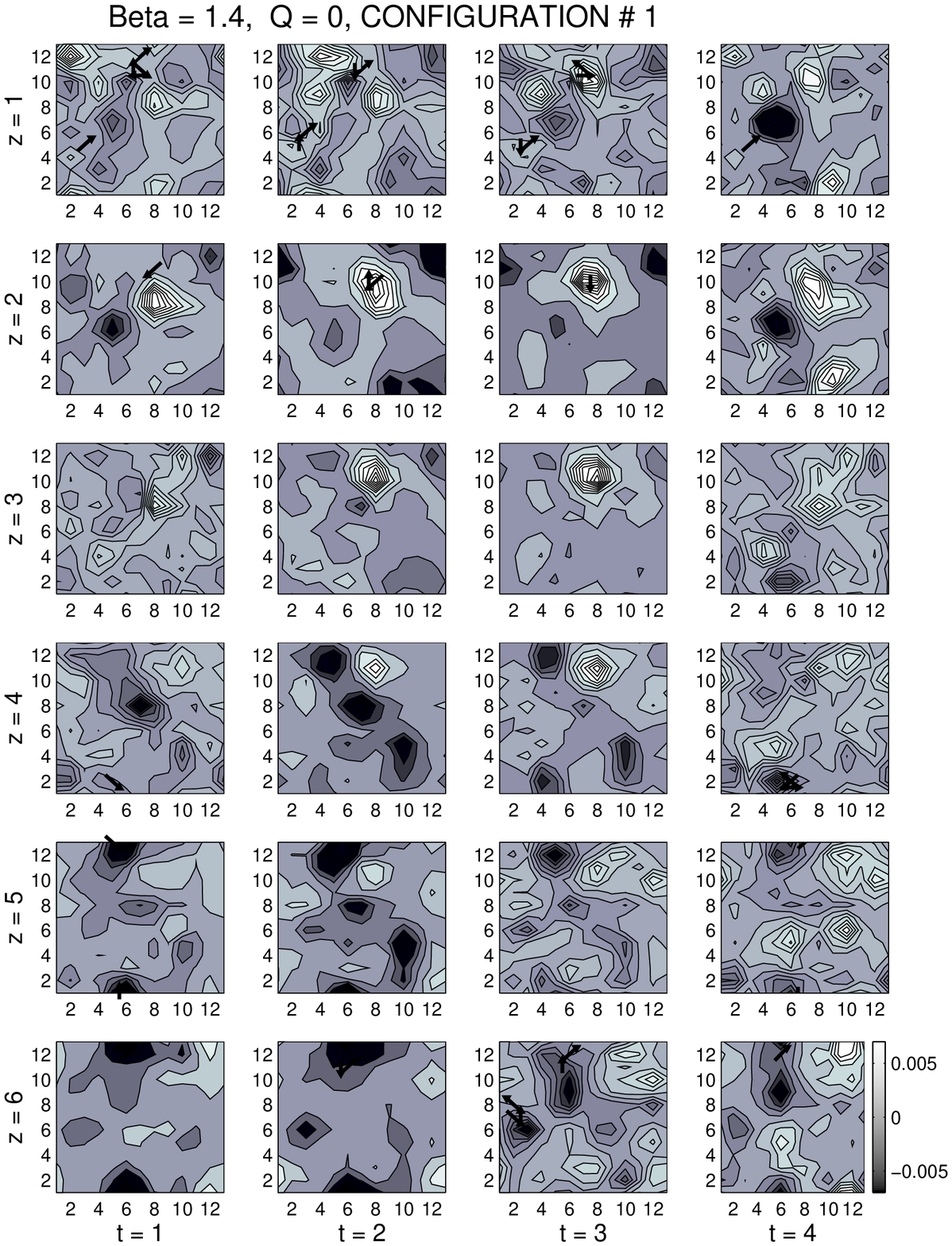}
\\
\\
{FIG. 4a }
\end{tabular}
\end{figure}

\newpage

\begin{figure}[!thb]
\label{fig:Q-2beta1_2}
\begin{tabular}{c}
\epsfxsize=16.0cm\epsffile{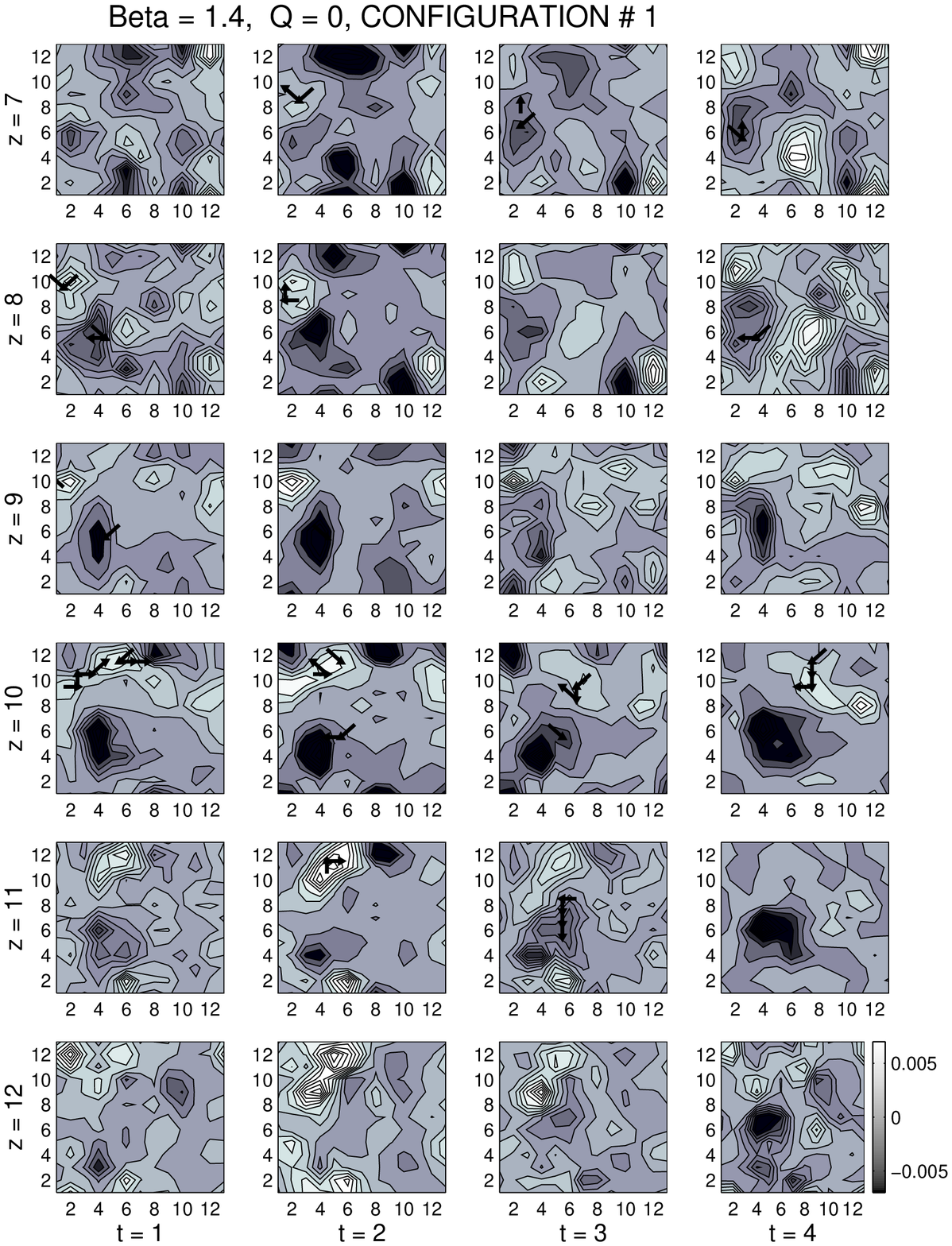}
\\
\\
{FIG. 4b }
\end{tabular}
\end{figure}

\newpage

\begin{figure}[!thb]
\label{fig:Q2_beta14_1}
\begin{tabular}{c}
\epsfxsize=16.0cm\epsffile{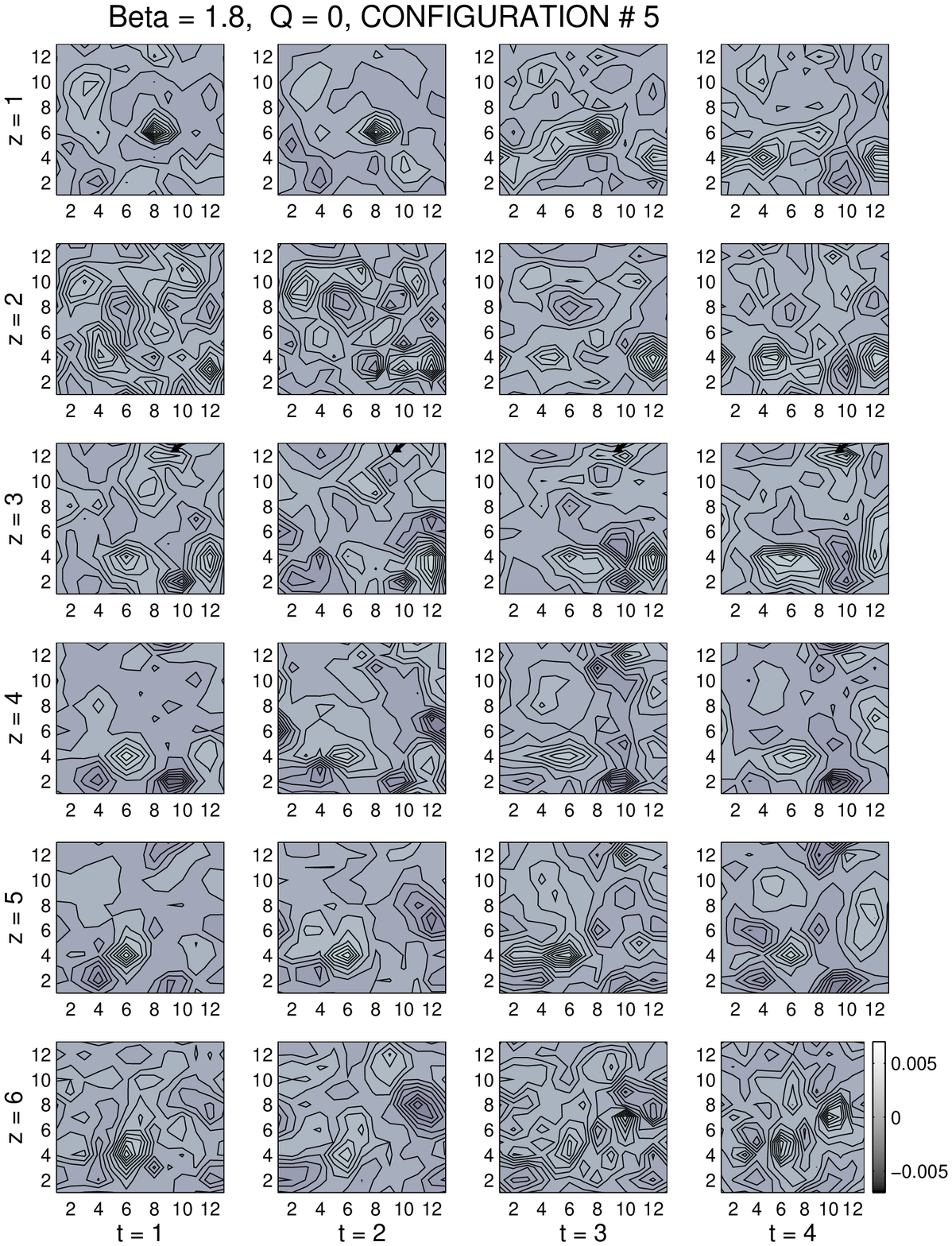}
\\
\\
{FIG. 4c }
\end{tabular}
\end{figure}

\newpage

\begin{figure}[!thb]
\label{fig:Q2_beta14_2}
\begin{tabular}{c}
\epsfxsize=16.0cm\epsffile{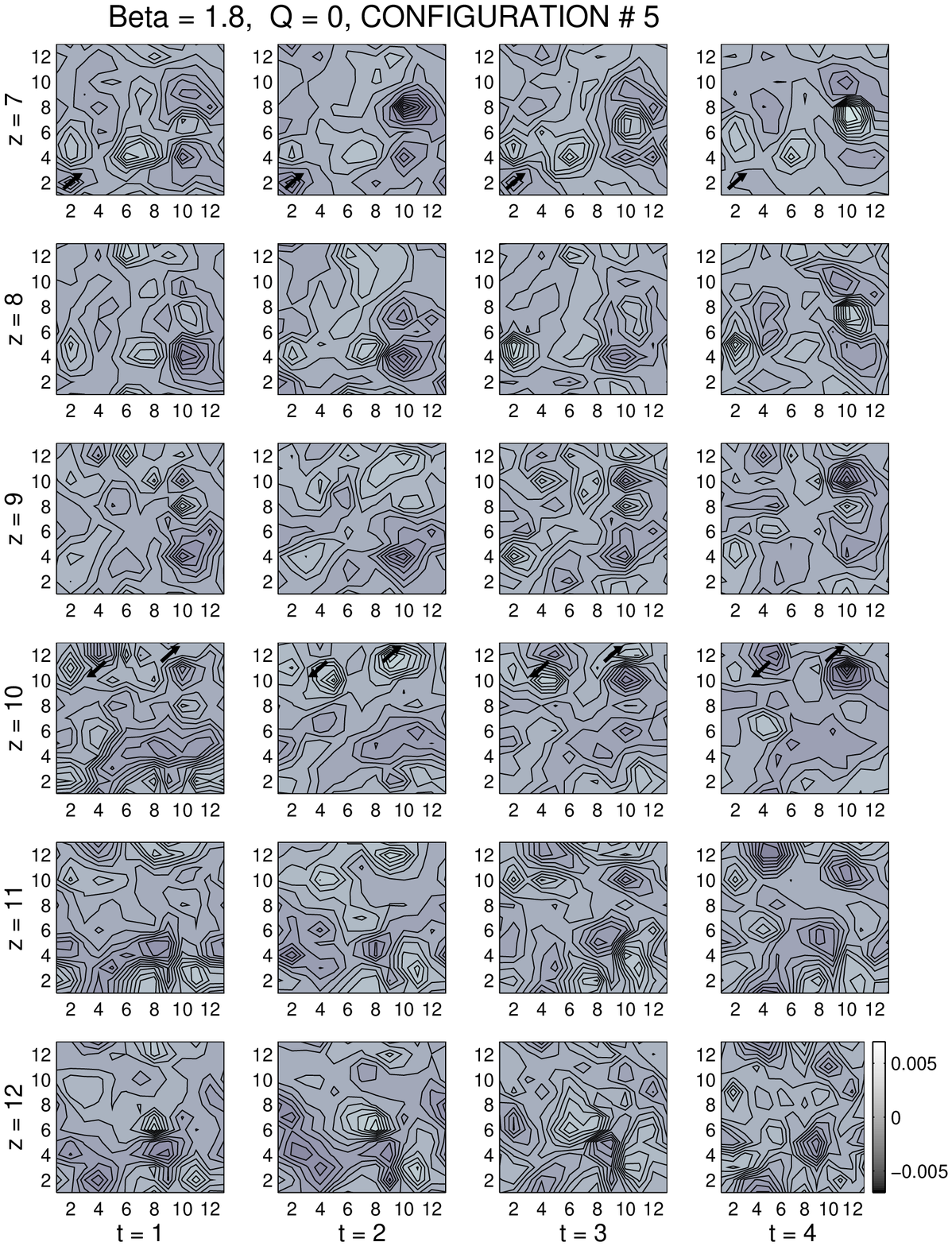}
\\
\\
{FIG. 4d }
\end{tabular}
\end{figure}

\newpage

\begin{figure}[!thb]
\label{fig:joint_s_and_q}
\begin{tabular}{c}
\\
\\
\\
\epsfxsize=15.0cm\epsffile{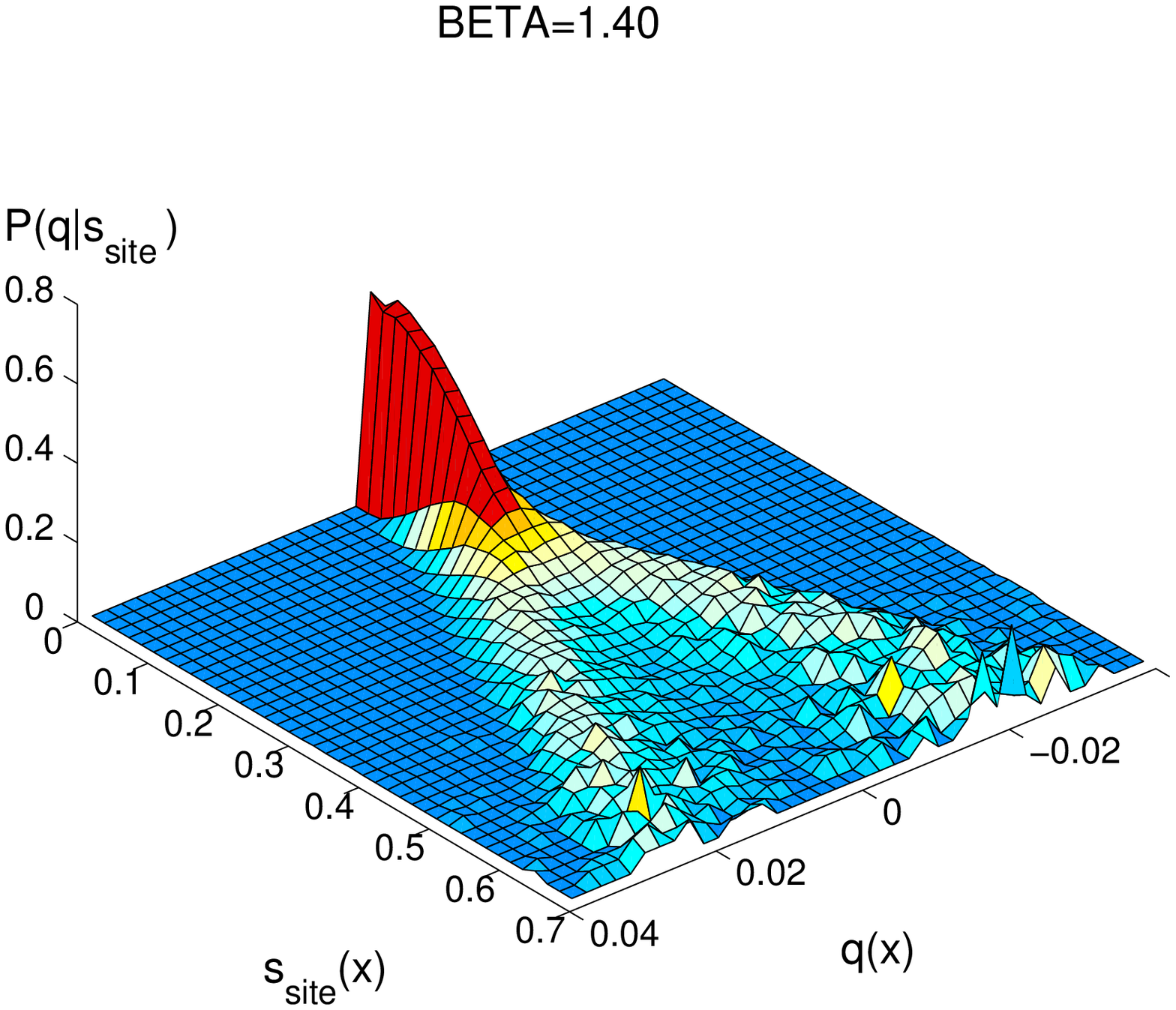}\\
\\
\\
\\
\\
\\
\\
\\
{FIG. 5 }
\end{tabular}
\end{figure}

\newpage

\begin{figure}[!thb]
\label{fig:avm_vs_action}
\begin{tabular}{c}
\hspace{2.1cm}\epsfxsize=11.0cm\epsffile{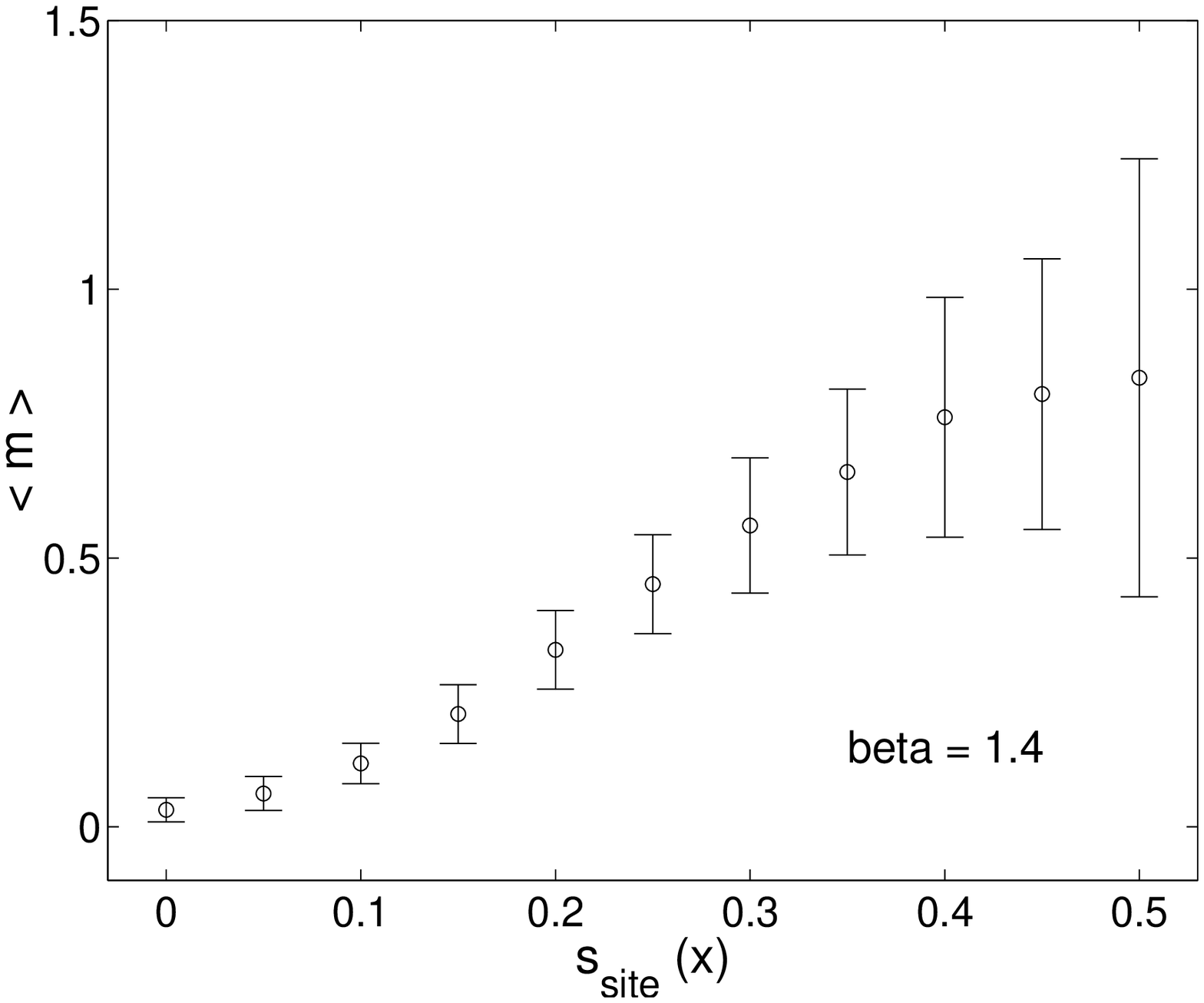}\\
\\
\hspace{1.3cm}\epsfxsize=10.3cm\epsffile{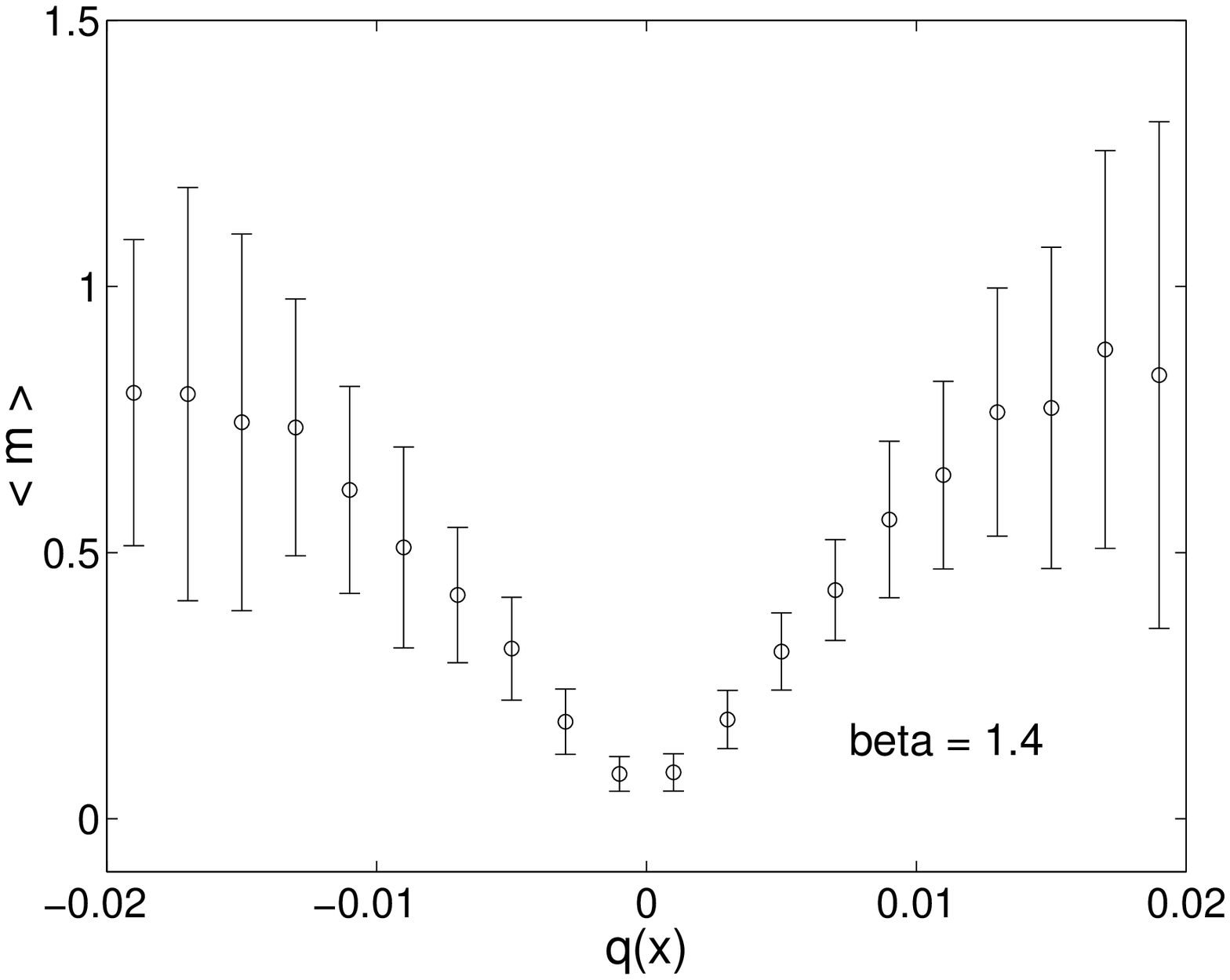}\\
\\
\\
\\
\\
\\
{FIG. 6 }
\end{tabular}
\end{figure}

\newpage

\begin{figure}[!thb]
\label{fig:hist_act_cube}
\begin{center}
\begin{tabular}{cr}
\rotate[l]{{ \large Action, beta=1.8} } &
\rotate[l]{ \epsfxsize=10.0cm\epsfysize=15.0cm\epsffile{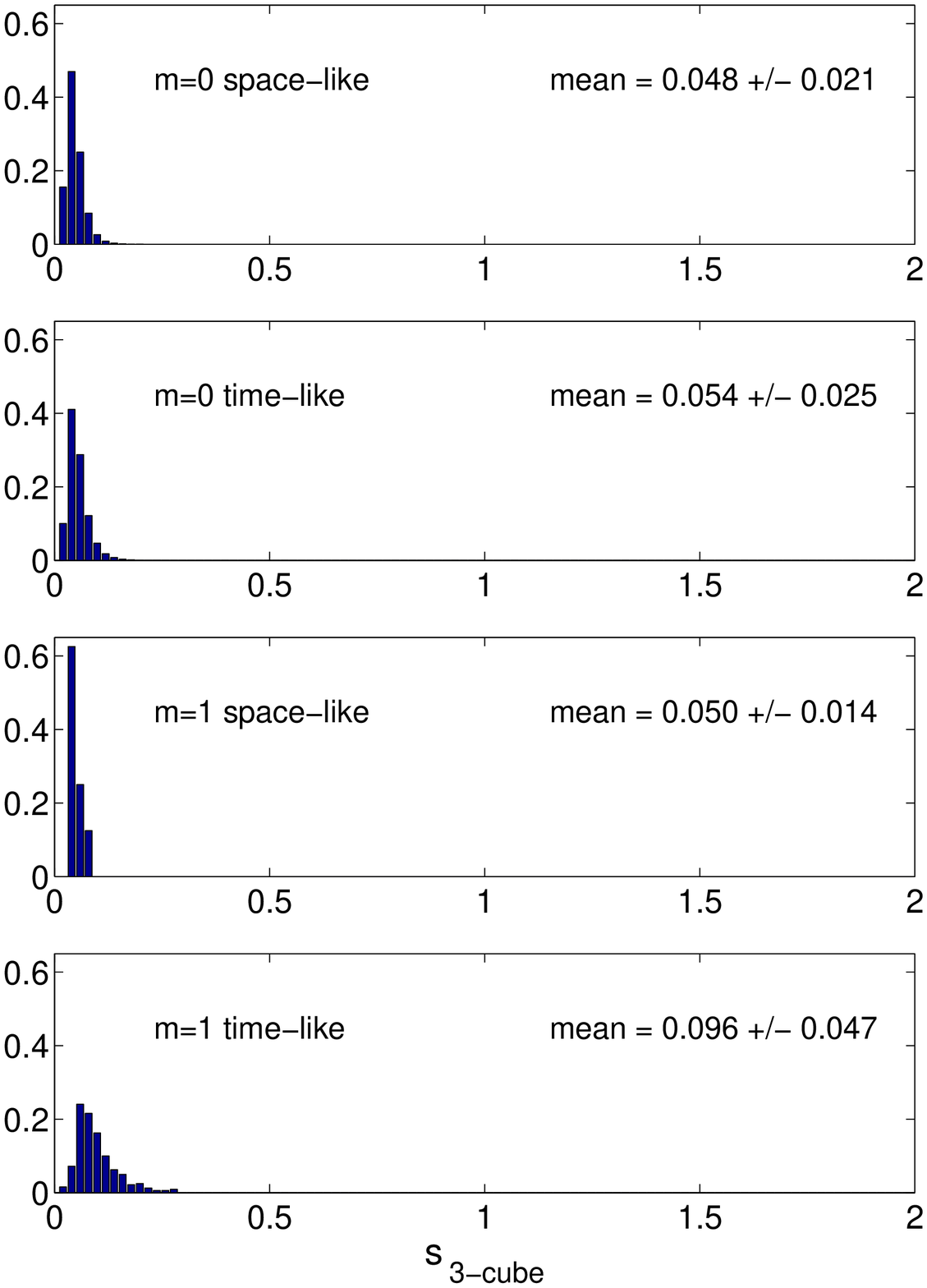}} \\
\\
\rotate[l]{{ \large Action, beta=1.4} } & 
\rotate[l]{ \epsfxsize=10.0cm\epsfysize=15.0cm\epsffile{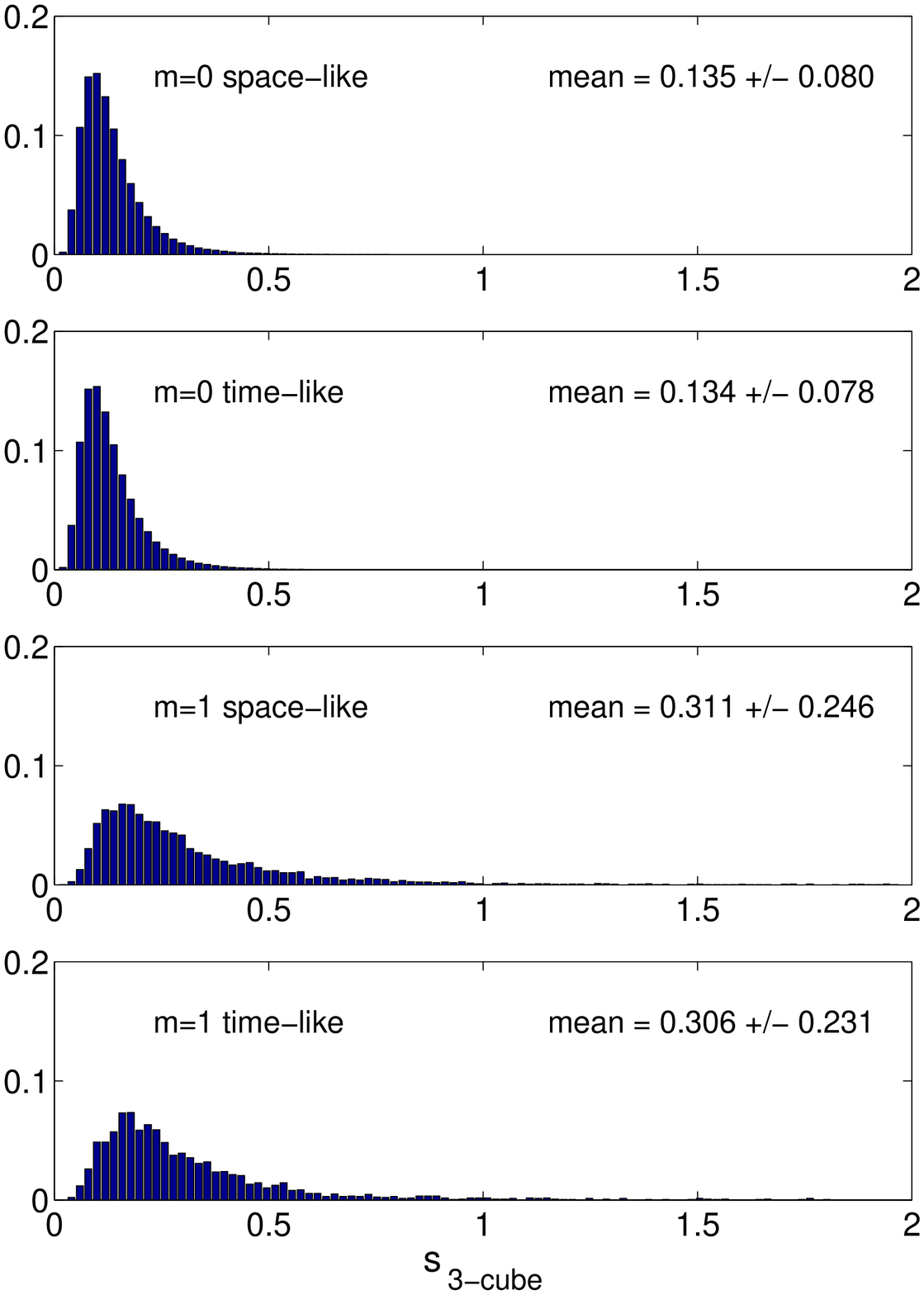}} \\
\\
\end{tabular}
{FIG. 7 }
\end{center}
\end{figure}

\newpage

\begin{figure}[!thb]
\label{fig:hist_sum_charge}
\begin{center}
\begin{tabular}{cr}
\rotate[l]{{ \large Topological Charge, beta=1.8} } & 
\rotate[l]{ \epsfxsize=10.0cm\epsfysize=15.0cm\epsffile{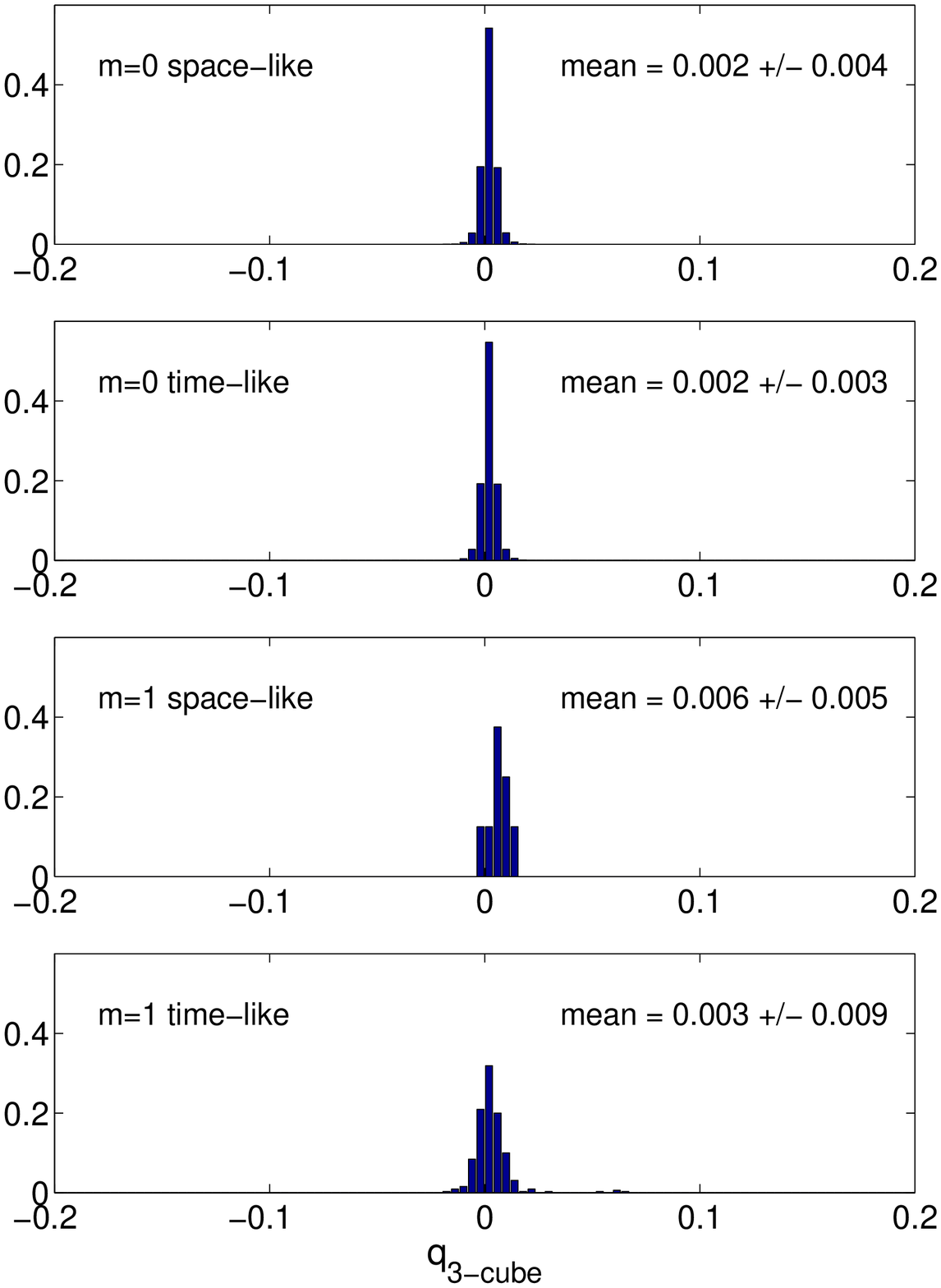}} \\
\\
\rotate[l]{{ \large Topological Charge, beta=1.4} } & 
\rotate[l]{ \epsfxsize=10.0cm\epsfysize=15.0cm\epsffile{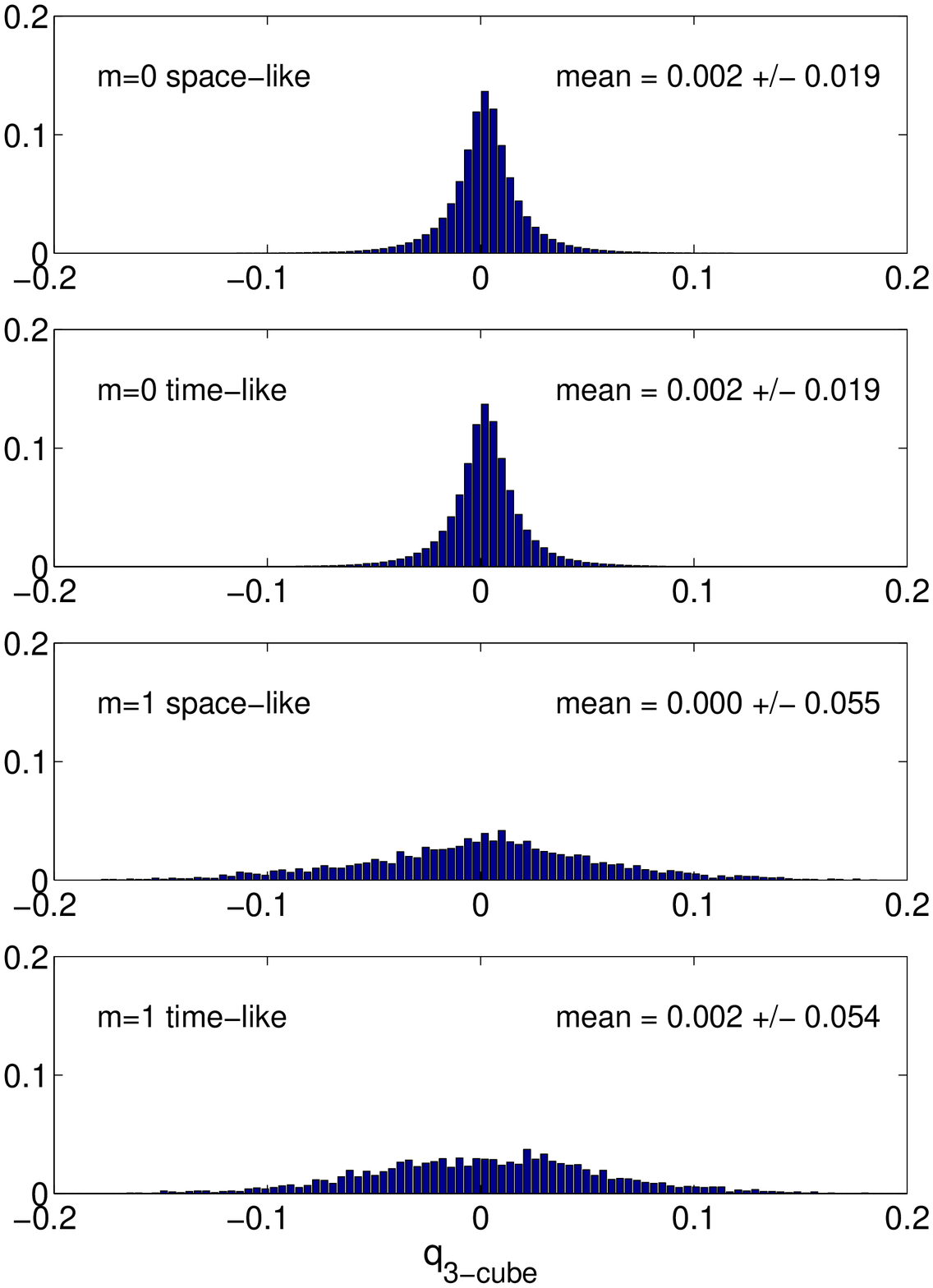}} \\
\\
\end{tabular}
{FIG. 8 }
\end{center}
\end{figure}

\newpage

\begin{figure}[!thb]
\label{fig:corr_unsmooth}
\begin{center}
\begin{tabular}{cr}
\rotate[l]{{ \large Topological Charge, beta=1.4} } & 
\rotate[l]{ \epsfxsize=10.0cm\epsfysize=15.0cm\epsffile{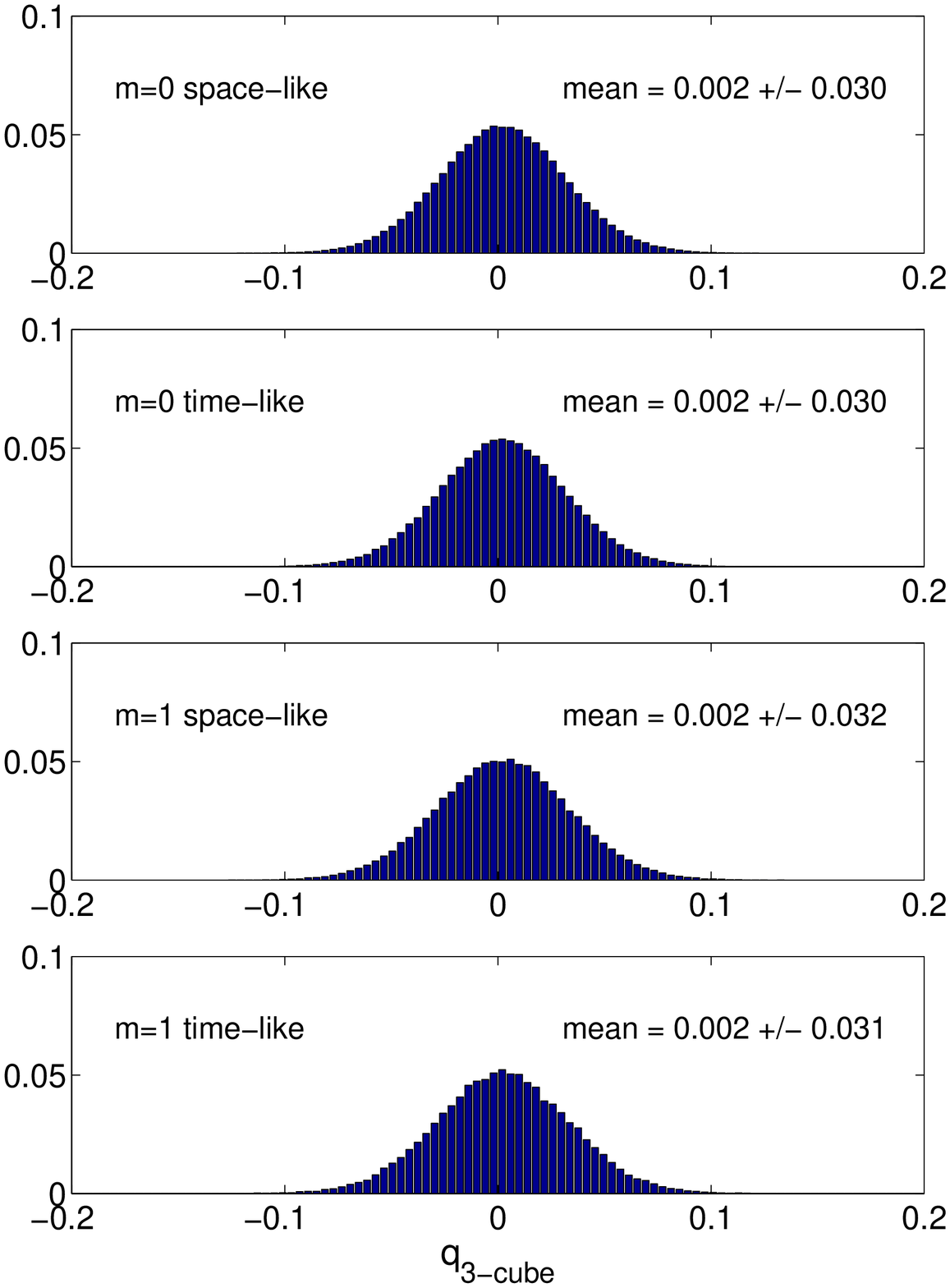}}  \\
\\
\rotate[l]{{ \large Action, beta=1.4} } & 
\rotate[l]{ \epsfxsize=10.0cm\epsfysize=15.0cm\epsffile{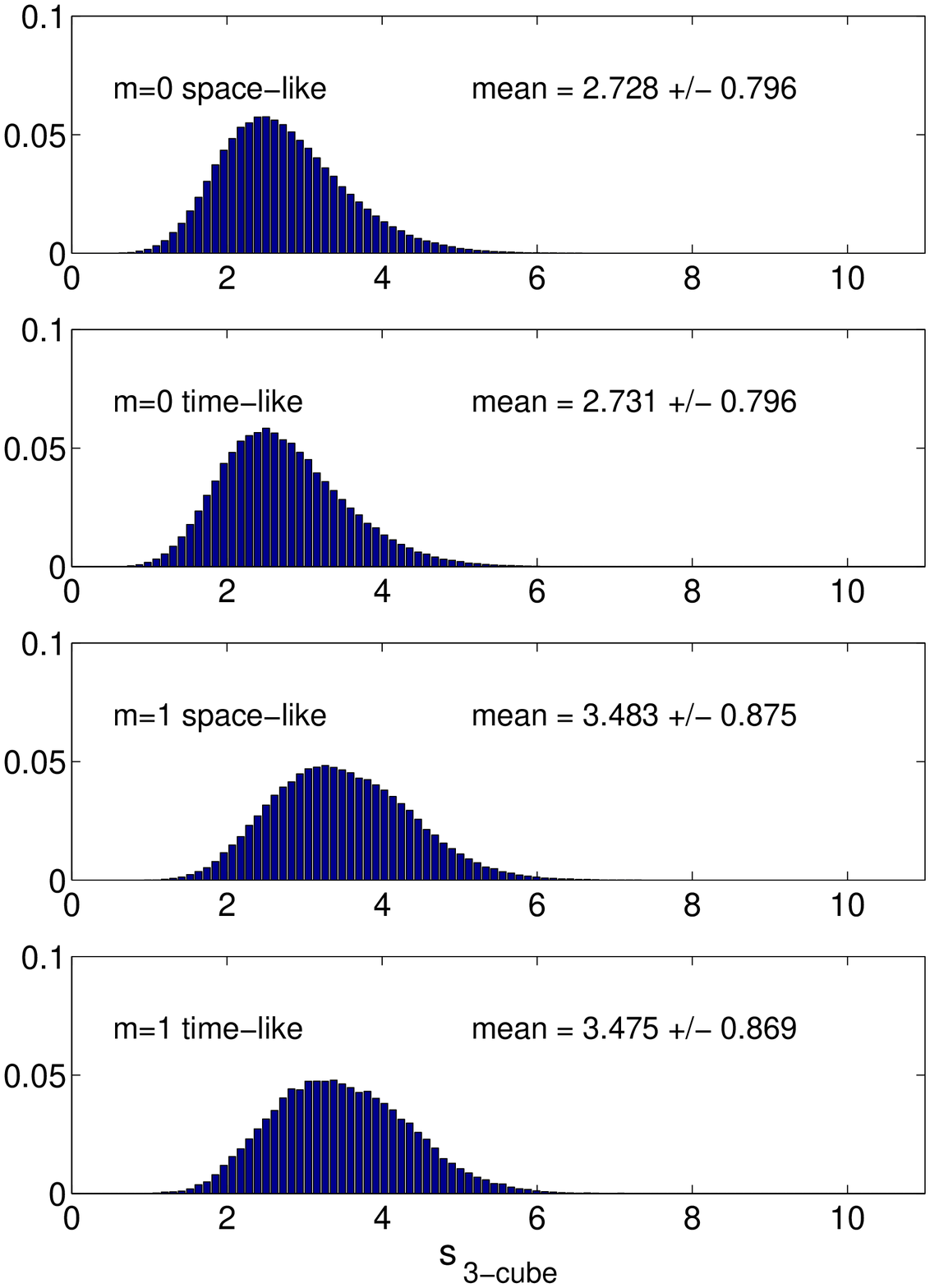}} \\
\\
\end{tabular}
{FIG. 9 }
\end{center}
\end{figure}

\newpage

\begin{figure}[!thb]
\label{fig:s_q_per_length_100}
\begin{tabular}{c}
{ \large  beta=1.4 } \\
\\
\epsfxsize=15.0cm\epsffile{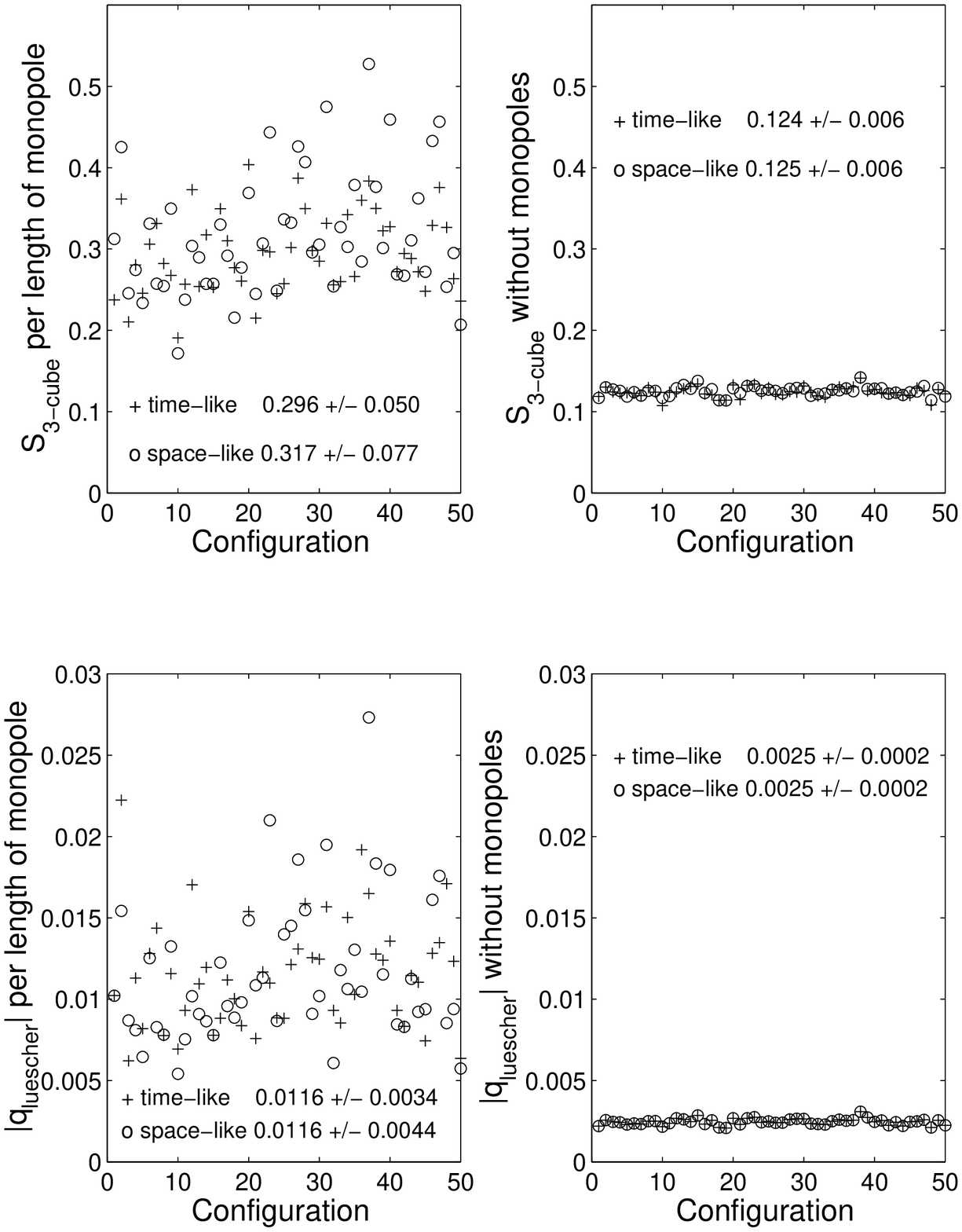}
\\
{FIG. 10}
\end{tabular}
\end{figure}

\newpage

\begin{figure}[!thb]
\label{fig:s_q_per_length_140}
\begin{tabular}{c}
{ \large  beta=1.8 } \\
\\
\epsfxsize=15.0cm\epsffile{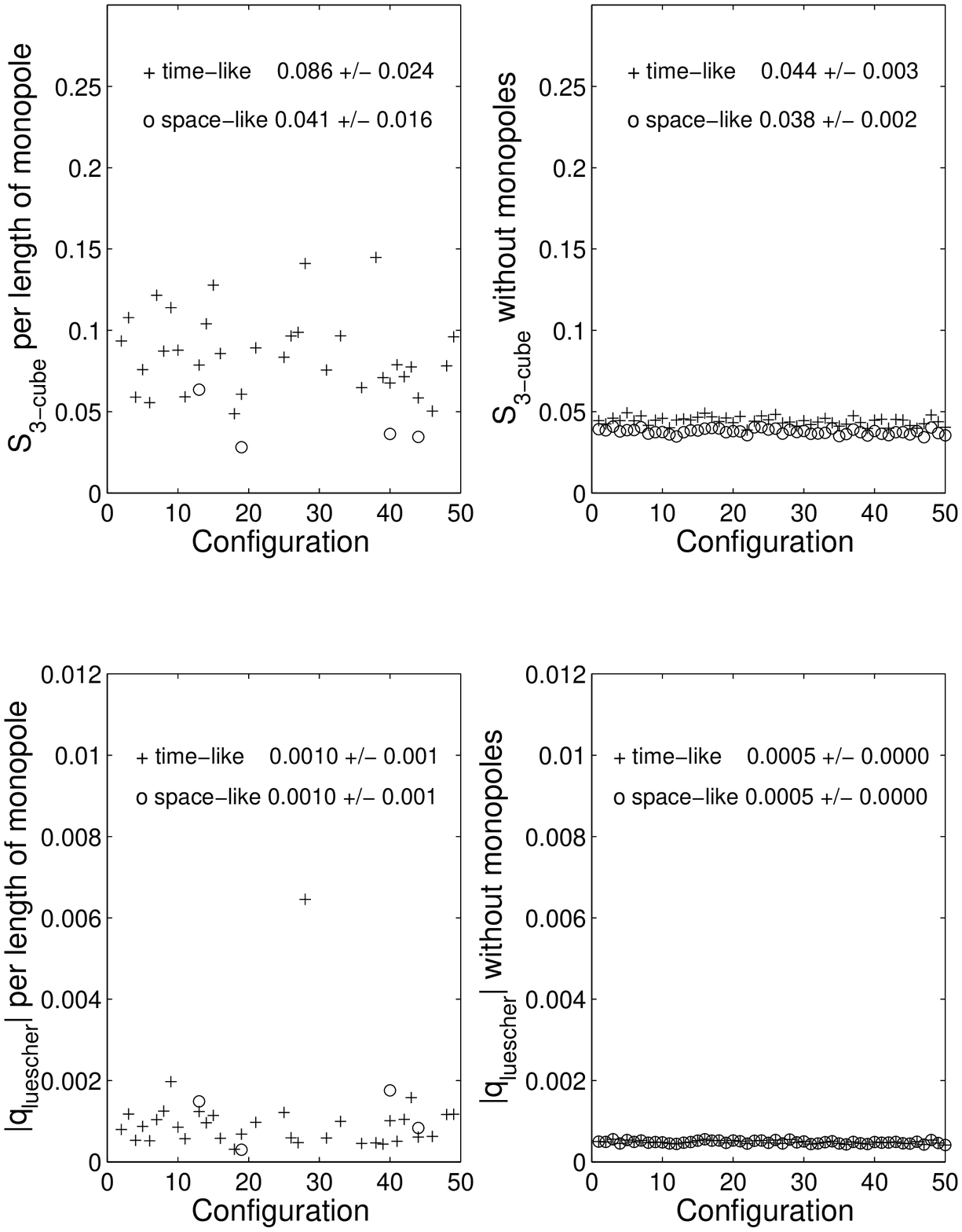}
\\
{FIG. 11}
\end{tabular}
\end{figure}


\begin{figure}[!thb]
\label{fig:mon_length}
\begin{tabular}{c}
\\
\\
\epsfxsize=16.0cm\epsfxsize=10.0cm\epsffile{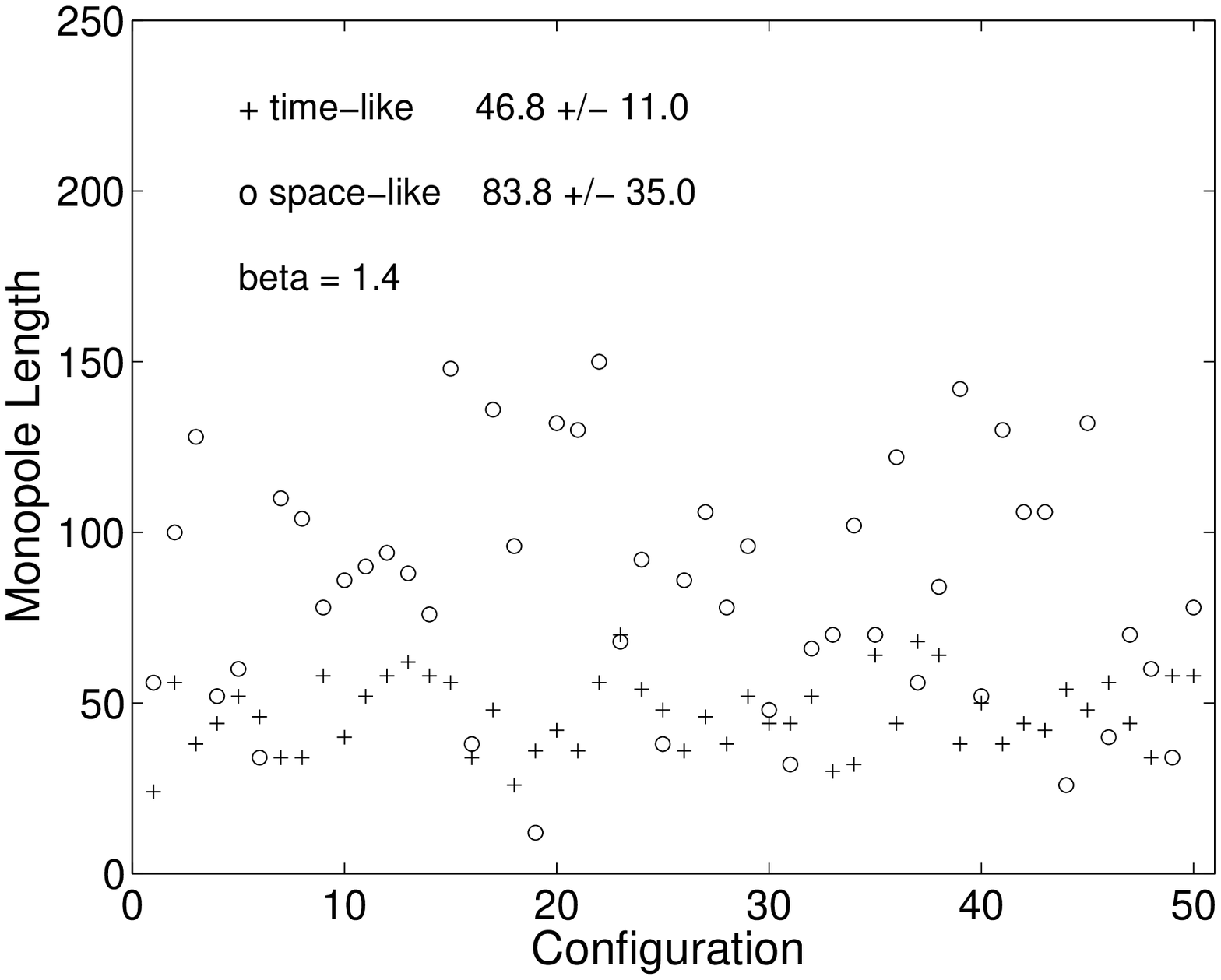}
\\
\\
\epsfxsize=16.0cm\epsfxsize=10.0cm\epsffile{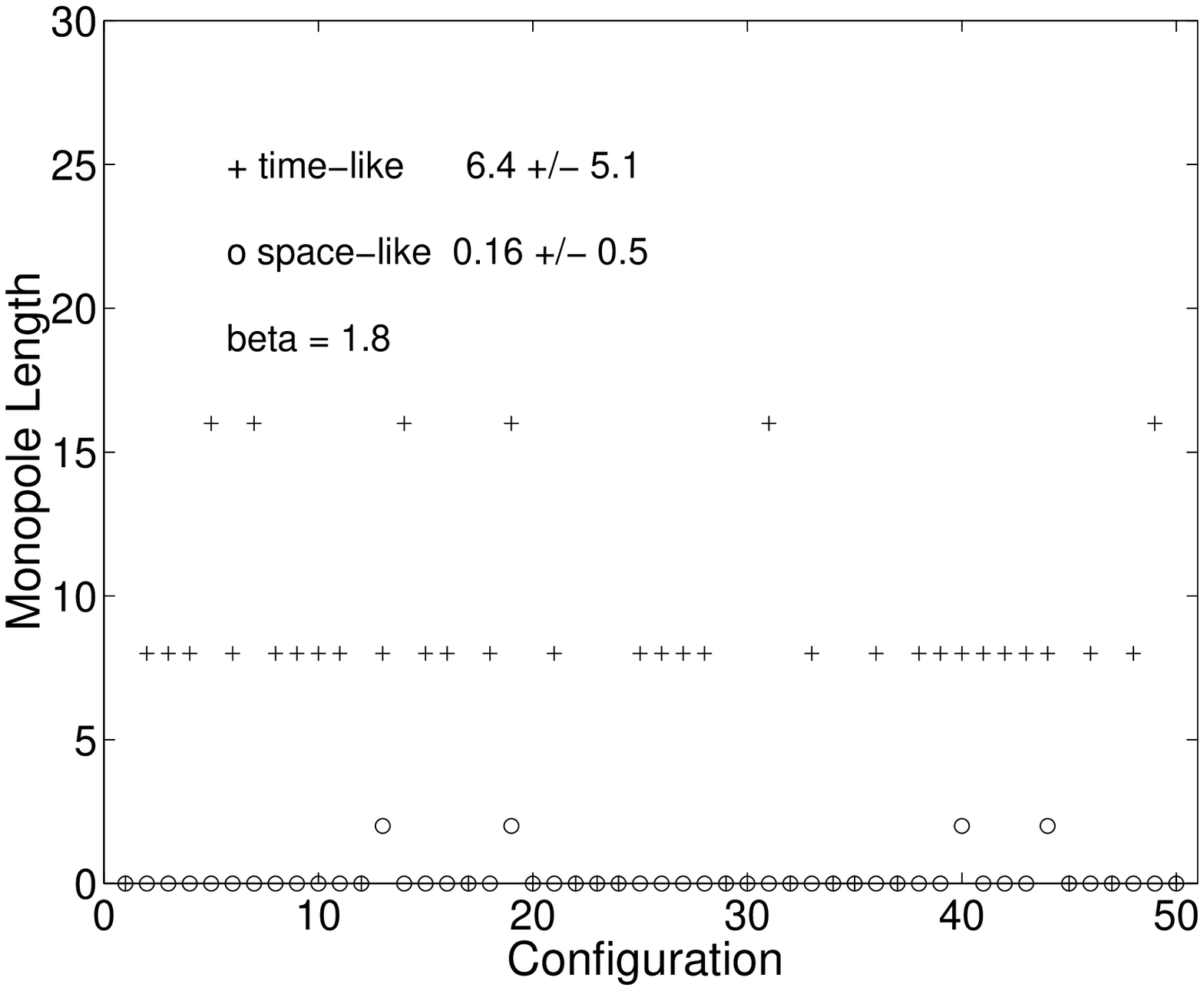}
\\
\\
\\
\\
\\
\\
{FIG. 12}
\end{tabular}
\end{figure}

\end{document}